\documentclass[a4paper,11pt]{article}
\pdfoutput=1 
\usepackage{jheppub} 
\usepackage[T1]{fontenc} 
\usepackage{epsfig}
\usepackage{subcaption}
\usepackage{color}
\usepackage{float}
\usepackage{slashed}
\usepackage{accents}
\def\beq{\begin{equation}}
\def\eeq{\end{equation}}
\def\beqn{\begin{eqnarray}}
\def\eeqn{\end{eqnarray}}

\def\N{{\cal N}}

\def\met{E_T \hspace*{-1.1em}/\hspace*{0.5em}}
\def\fb{\, {\rm fb}}

\def\tev{\, \, {\rm TeV}}
\def\gev{\, \, {\rm GeV}}
\def\neu{\tilde{\chi}_1^0}

\def\tla{\tilde{\tau}_1}
\def\tlb{\tilde{\tau}_2}

\def\smuL{\tilde{\mu}_L}

\def\C1{\tilde{\chi}_1^\pm}
\def\N2{\tilde{\chi}_2^0}
\def\gm2{{(g-2)}_\mu}

\title{Closing in on the chargino contribution to the muon $g-2$ in the MSSM:  current LHC constraints}

\author[a,b]{Kaoru Hagiwara,}
\author[a,c]{Kai Ma}
\author[b]{and Satyanarayan Mukhopadhyay}

\affiliation[a]{KEK Theory Center, Tsukuba 305-0801, Japan}
\affiliation[b]{PITT-PACC, Department of Physics and Astronomy, University of Pittsburgh, PA 15260, USA}
\affiliation[c]{Department of Physics, Shaanxi Sci-Tech University, Hanzhong 723000, Shaanxi, China}

\emailAdd{kaoru.hagiwara@kek.jp}
\emailAdd{makainca@yeah.net}
\emailAdd{satya@pitt.edu}

\preprint{PITT-PACC 1621}
\abstract{We revisit the current LHC constraints on the electroweak-ino sector parameters in the minimal supersymmetric standard model (MSSM) that are relevant to explaining the $(g-2)_\mu$ anomaly via the dominant chargino and muon sneutrino loop. Since the LHC bounds on electroweak-inos become weaker if they decay via an intermediate stau or a tau sneutrino instead of the first two generation sleptons, we perform a detailed analysis of the scenario with a bino as the lightest supersymmetric particle (LSP) and a light stau as the next-to-lightest one (NLSP).  Even in this scenario, the chargino sector parameters in the MSSM that can account for the $(g-2)_\mu$ anomaly within $1\sigma$ are already found to be significantly constrained by the 8 TeV LHC and the available subset of the 13 TeV LHC limits. We also estimate the current LHC exclusions in the left-smuon (and/or left-selectron) NLSP scenario from multilepton searches, and further combine the constraints from the multi-tau and multi-lepton channels for a mass spectrum in which all three generations of sleptons are lighter than the chargino. In the latter two cases, small corners of the $1 \sigma$ favoured region for $(g-2)_\mu$ are still allowed at present.}

\begin{document} 
\maketitle
\flushbottom
\section{Introduction}
The anomalous magnetic moment of the muon, $a_\mu = (g-2)_\mu/2$, is an accurately measured quantity, which, at the same time, is precisely predicted within the standard model (SM)~\cite{Review}. Consequently, it is an important testing ground for the SM, since new states at the electroweak scale with coupling to muons can potentially contribute to $a_\mu$ via quantum corrections. Due to the chirality-flip nature of the magnetic moment operator, the sensitivity of $a_\mu$ to new particles coupling to leptons is parametrically higher by a factor of $(m_\mu/m_e)^2$, than that of the electron anomalous magnetic moment ($a_e$), despite the latter's higher precision measurements. 

The BNL E821 experiment~\cite{Bennett,Roberts:2010cj} provides the current best measurement for $a_\mu$, which reads
\begin{equation}
a_{\mu}^{\rm EXP} = (11~659~208.9 \pm 6.3) \times 10^{-10}.
\end{equation}
In the SM, $a_\mu$ receives relevant  QED, electroweak and hadronic contributions, with the hadronic contribution being the dominant source of the uncertainty in the theoretical prediction. We refer the reader to Ref.~\cite{Hagiwara_gm2SM} for a detailed recent discussion of the different contributions and their associated uncertainty estimates, while for the current study, we adopt the following SM prediction for $a_\mu$~\cite{Hagiwara_gm2SM}:
\begin{equation}
a_{\mu}^{\rm SM} = (11~659~182.8 \pm 4.9) \times 10^{-10}.
\end{equation}
Thus, the measured value of  $a_\mu$ is larger than the SM prediction by
\begin{equation}
a_{\mu}^{\rm EXP} - a_{\mu}^{\rm SM} = (26.1 \pm 8.0) \times 10^{-10},
\end{equation}
which corresponds to a $3.3 \sigma$ discrepancy. 

The measurement uncertainty on $a_\mu$ is expected to be reduced further by two upcoming experiments. The FermiLab FNAL E989 experiment~\cite{Grange:2015fou}, due to start data taking in 2017, is projected to achieve a factor of four reduction in the current measurement uncertainty. Using a completely different technology with an ultra-cold muon beam which does not share the same systematic uncertainties associated with the BNL and FNAL experiments, the J-PARC E34 experiment also has a competitive potential~\cite{JPARC}. Therefore, if the current $3.3\sigma$ deviation is truly a sign of physics beyond the SM, these follow-up measurements would lead to an enhanced statistical significance for the present discrepancy. Simultaneously, the error in the theoretical prediction within the SM also needs better control, with the recent progress in estimating the hadronic light-by-light contribution from lattice QCD computations~\cite{lattice} being promising in this regard.

The minimal supersymmetric standard model (MSSM) contains the necessary ingredients to accommodate the above discrepancy through contributions from scalar muons, the muon sneutrino, charginos and neutralinos (the latter two will be referred to as electroweak-inos in the following). A number of studies have been devoted to the computation of the MSSM contribution to the $\gm2$~\cite{Moroi:1995yh, Chattopadhyay:1995ae, Lopez:1993vi, Cho:2000sf, Cho:2001nfa, Heinemeyer:2003dq, Stockinger:2006zn, Marchetti:2008hw, vonWeitershausen:2010zr, Cho_gm2, Fargnoli:2013zia}, the constraints on the relevant parameter space from a subset of the 8 TeV LHC data~\cite{gm2_LHC}, as well as the role of future $e^+e^-$ colliders (such as the ILC) in probing the light electroweak MSSM sector that can contribute to the $\gm2$~\cite{gm2_ILC}. The goal of the present study is to revisit the current constraints on the dominant contribution to $\gm2$ in the MSSM, namely that of the chargino and muon-sneutrino loop, in the light of recent data from both the 8 and 13 TeV runs of the LHC.

The ATLAS and CMS collaborations have carried out multiple sets of analyses in their search for the electroweak sector of the MSSM, and have practically covered all possible decay modes of the lighter chargino ($\C1$) and the second lightest neutralino ($\N2$). The search for electroweak-inos at the LHC is complex due to the multitude of decay modes available, determined by the mass hierarchies and mixing angles in the chargino, neutralino and slepton sectors. Due to the complexity in interpreting the searches in a multi-dimensional parameter space, so far, most of the ATLAS and CMS results assume one particular decay mode of the $\C1$ and $\N2$ to have a $100\%$ branching ratio (BR), and interpret the results within a simplified model setting. Implications of the LHC search results for the electroweak MSSM sector have also been explored in a number of phenomenological studies~\cite{ewino_LHC}.

In the most general scenario, the MSSM parameters relevant for the electroweak searches of our interest in this study are the mass parameters for the bino ($M_1$), wino ($M_2$) and Higgsino ($\mu$), the masses of the lighter ($M_{\tilde{\tau}_1}$) and heavier stau ($M_{\tilde{\tau}_2}$), the mixing angle in the stau sector ($\theta_{\tilde{\tau}}$), the soft masses of the left-smuon ($M_{\smuL}$) and left-selectron ($M_{\tilde{e}_L}$), and the ratio of the vacuum expectation values of the two Higgs doublets ($\tan \beta$).  All of these parameters enter the search for electroweak-inos and the left-smuon, and all except the stau and selectron sector parameters also determine the MSSM contribution to the $\gm2$ at the one-loop level. Therefore, in order to perform a completely general analysis of the current status of $\gm2$ within the MSSM in the light of LHC searches, we need to perform a nine parameter global analysis, including, at the same time, several relevant LHC search channels. The computational resources required for this analysis is beyond the scope of our present study. However, as we shall argue in the subsequent sections, a relatively small set of computations is sufficient to obtain a broad understanding of the current constraints.

The rest of the sections and the general strategy adopted in this study are as follows. In Sec.~\ref{sec:gm2}, we provide a brief overview of the different contributions to the $\gm2$ in the MSSM. In Sec.~\ref{smuon} we determine, to what extent the current LHC bound on the left-smuon mass parameter can be modified in the presence of $\C1$ and $\N2$ states lighter than the $\smuL$. Such a lower bound, in turn, leads to an estimate of the maximum possible chargino mass that can explain the $\gm2$ anomaly within $1\sigma$, depending upon the relevant mixing matrices in the chargino and neutralino sectors. 

As far as the $\C1$ and $\N2$ searches are concerned, we first enumerate in Sec.~\ref{sec:chargino}  the different possible mass hierarchies between the $\C1/\N2$ and the sleptons that determine the most relevant search channel(s). We then proceed to obtain the current LHC constraints on the stau NLSP scenario in Sec.~\ref{stau}, using the ATLAS search results in the $\geq 2\tau_h+\met$ channel. Translating the LHC simplified model-based search constraints in this case is somewhat involved, due to the required modelling of the tau jet identification and reconstruction efficiencies. Therefore, we perform a detailed Monte Carlo (MC) analysis, including detector resolution effects, to recast the 8 and 13 TeV LHC search limits in the $\geq 2\tau_h+\met$ channel to the constraints in the $\mu-M_2$ plane. Variation in the LHC bounds due to changes in the LSP mass and the stau mixing angle are also illustrated here. 

In Sec.~\ref{trilepton}, we consider the smuon/selectron NLSP scenario, and utilize the LHC searches in the trilepton and $\met$ channel to estimate the current LHC limits on the electroweak-ino sector parameters for this mass hierarchy. In Sec.~\ref{sec:sc3}, we then combine the limits from the multi-tau search channel and the trilpeton channel to obtain the LHC constraints for the scenario in which all three generation sleptons are lighter than the chargino. In Sec.~\ref{combi}, we revisit the $1\sigma$ favoured region for the $\gm2$, in the light of the above constraints, paying particular attention to the mass hierarchies. We provide a summary of our results in Sec.~\ref{sum}. Our MC simulation setup for the $\geq 2\tau_h+\met$ search and its validation against the ATLAS 8 and 13 TeV search results are discussed in the Appendix.

\section{Muon $(g-2)$ in the MSSM -- a brief overview}
\label{sec:gm2}
In the MSSM, $a_{\mu}$ receives contributions at the one loop level from the chargino $(\tilde{\chi}^{\pm})-$muon sneutrino $(\tilde{\nu}_{\mu})$ loop and the neutralino $(\tilde{\chi}^0)-$smuon $(\tilde{\mu})$ loop~\cite{Moroi:1995yh, Chattopadhyay:1995ae, Lopez:1993vi, Cho:2000sf, Cho:2001nfa, Heinemeyer:2003dq, Stockinger:2006zn, Marchetti:2008hw, vonWeitershausen:2010zr, Cho_gm2, Fargnoli:2013zia}. In order to understand the dependence of the supersymmetric contribution to the $\gm2$ on different MSSM parameters, we follow the discussion in Ref.~\cite{Cho_gm2}, and classify the contributions in the weak eigenstate basis. In this basis, 
the leading terms in the $m_{\mathrm{EW}}/m_{\mathrm{SUSY}}$ expansion (here, $m_{\mathrm{EW}}$ stands for $m_{\mu}$, $m_W$ or $m_Z$ and $m_{\mathrm{SUSY}}$ stands for $m_{\tilde{\mu}}$, $m_{\tilde{\nu}}$, $M_1$, $M_2$ or $\mu$), are given by five one-loop diagrams, see Fig.~1 and Eq.~(2.6) in Ref.~\cite{Cho_gm2}. The dominant contributions proportional to the wino-Higgsino mixing are given by the charged wino -- charged Higgsino -- muon sneutrino loop and the neutral wino -- neutral Higgsino -- left smuon loop as follows:

\begin{subequations}\label{eq:gm21}
\begin{align}
\Delta a_{\mu}(\tilde{W}-\tilde{H}, \tilde{\nu}_{\mu})&=\frac{g^2m_{\mu}^2}{8\pi^2}\frac{M_2\mu\tan\beta}{m_{\tilde{\nu}_{\mu}}^4}
F_a(M_2^2/m_{\tilde{\nu}_{\mu}}^2,\mu^2/m_{\tilde{\nu}_{\mu}}^2),\label{a}\\
\Delta a_{\mu}(\tilde{W}-\tilde{H}, \tilde{\mu}_L)&=-\frac{g^2m_{\mu}^2}{16\pi^2}\frac{M_2\mu\tan\beta}{m_{\tilde{\mu}_L}^4}
F_b(M_2^2/m_{\tilde{\mu}_L}^2,\mu^2/m_{\tilde{\mu}_L}^2)\label{d},
\end{align}
\end{subequations}
where, the loop functions in the above expressions are given as
 \begin{subequations}
\begin{align}
F_a(x,y) &= -\frac{G_3(x) - G_3(y)}{x-y},\\
F_b(x,y) &= -\frac{G_4(x) - G_4(y)}{x-y},\\
G_3(x) &= \frac{1}{2(x-1)^3}[(x-1)(x-3)+2\ln{x}],\\
G_4(x) &=\frac{1}{2(x-1)^3}[(x-1)(x+1)-2x\ln{x}].
\end{align}
\end{subequations}
$F_a(x,y)$ and $F_b(x,y)$ are defined to be positive for all positive $x$ and $y$, and $F_a(x,y)$ is always larger than $F_b(x,y)$ for the same arguments~\cite{Cho_gm2}.
These two contributions are enhanced when the wino and Higgsino are maximally mixed, which requires $|M_2/\mu| \sim \mathcal{O}(1)$. In the approximation of equal masses for the left-smuon and the muon sneutrino, the arguments of $F_{a,b}(x,y)$ in Eqs.~(\ref{a}, \ref{d}) are the same. Taking into account the difference of a factor of two, and the fact that $F_a(x,y) > F_b(x,y)$, we can infer that the positive discrepancy between the data and the SM prediction can be explained with $M_2\mu > 0$. We note that the right-smuon is not relevant for these two contributions. 

The terms proportional to the bino-Higgsino mixing are given by the bino -- neutral Higgsino -- left-smuon loop and the neutral Higgsino -- bino -- right-smuon loop as follows:
\begin{subequations}\label{eq:gm22}
\begin{align}
\Delta a_{\mu}(\tilde{B}-\tilde{H},\tilde{\mu}_L)&=\frac{g_Y^2 m_{\mu}^2}{16\pi^2}\frac{M_1\mu\tan\beta}{m_{\tilde{\mu}_L}^4}
F_b(M_1^2/m_{\tilde{\mu}_L}^2,\mu^2/m_{\tilde{\mu}_L}^2),\label{c}\\
\Delta a_{\mu}(\tilde{B}-\tilde{H}, \tilde{\mu}_R)&=-\frac{g_Y^2 m_{\mu}^2}{8\pi^2}\frac{M_1\mu\tan\beta}{m_{\tilde{\mu}_R}^4}
F_b(M_1^2/m_{\tilde{\mu}_R}^2,\mu^2/m_{\tilde{\mu}_R}^2).\label{e}
\end{align}
\end{subequations}
In a scenario with a light bino-like LSP as considered here, these terms are generally much smaller than the ones in Eq.~\ref{eq:gm21}, due to the smaller hypercharge coupling (with $g_Y^2=g^2 \tan^2 \theta_W$, $g$ being the $\text{SU}(2)_L$ gauge coupling, and $\sin^2 \theta_W = 0.23$), as well as due to small bino-Higgsino mixing. Since the focus of our study is the dominant chargino-muon sneutrino loop contribution in Eq.~\ref{a}, which does not depend upon the right-smuon mass, for our discussion, we shall assume the right-smuon to be decoupled. This implies that the contribution from Eq.~\ref{e} will be negligible. The neutralino contribution in Eq.~\ref{c} is determined by the same set of parameters that enter Eq.~\ref{eq:gm21} as well as $M_1$, and it is positive for $M_1 \mu > 0$.

Finally, the contribution proportional to the left-smuon and right-smuon mixing is given by the bino -- left-smuon -- right-smuon loop and reads:
\begin{align}
\Delta a_{\mu}(\tilde{B},\tilde{\mu}_L-\tilde{\mu}_R)&=\frac{g_Y^2 m_{\mu}^2}{8\pi^2}\frac{\mu\tan\beta}{M_{1}^3}
F_b(m_{\tilde{\mu}_L}^2/M_1^2,m_{\tilde{\mu}_R}^2/M_1^2).\label{b}
\end{align}
This contribution can be enhanced for a large value of $\mu \tan \beta$, with a non-trivial dependence on $M_1$. The LHC analyses considered by us do not probe this term, which can be significant when both the left-smuon and the right-smuon are light, while the charginos can be much heavier. Once again, as for Eq.~\ref{e}, in the limit of a decoupled right-smuon, the contribution of Eq.~\ref{b} also becomes negligible. In our analysis, we include all five contributions given by Eqs.~\ref{eq:gm21}--\ref{b}. Although we have used the above expressions for $\Delta a_{\mu}$ in the weak eigenstate basis in the subsequent sections, we have checked that using the mass eigenstate basis expressions do not lead to any significant difference for the decoupled right-smuon scenario.

\section{Left-smuon mass limits from the LHC}
\label{smuon}
The ATLAS and CMS collaborations have searched for scalar muon pair production in the dilepton and missing transverse momentum ($\met$) channel, and with $20.3 \fb^{-1}$ of data from the 8 TeV LHC, ATLAS obtained the following $95\%$ C.L. lower bounds~\cite{ATLAS_Slepton}:
\begin{align}
M_{\tilde{\mu}^\pm_L} & >  300 \gev, \text{~with}~ \tilde{\mu}^\pm_R \text{~decoupled} \nonumber \\
M_{\tilde{\mu}^\pm_R} & >  230 \gev, \text{~with}~ \tilde{\mu}^\pm_L \text{~decoupled} \nonumber \\
M_{\tilde{\mu}^\pm_L,\tilde{\mu}^\pm_R} & >  320 \gev, \text{~for a common} ~\tilde{\mu}^\pm_{L,R}  ~\text{mass}.
\label{smuon_bound}
\end{align}
The quoted lower bounds correspond to the case where ${\rm BR}({\tilde{\mu}^\pm_{L,R}} \rightarrow \mu^\pm \neu) = 1$, and $M_{\neu}=0$ GeV. These limits are not very sensitive to the choice of $M_{\neu}$, unless ${\tilde{\mu}^\pm_{L,R}}$ and ${\neu}$ are nearly degenerate, and we find that the $M_{\tilde{\mu}^\pm_L}  >  300 \gev$ bound essentially remains unchanged for $M_{\neu}$ values of upto $150$ GeV~\cite{ATLAS_Slepton}. We note that although smuon masses lower than around $94 \gev$ are not probed by the ATLAS search, such lower mass regions are excluded by the LEP2 experiment~\cite{Abbiendi:2003ji}.
\begin{figure}[t]
\centering
\includegraphics[scale=0.4]{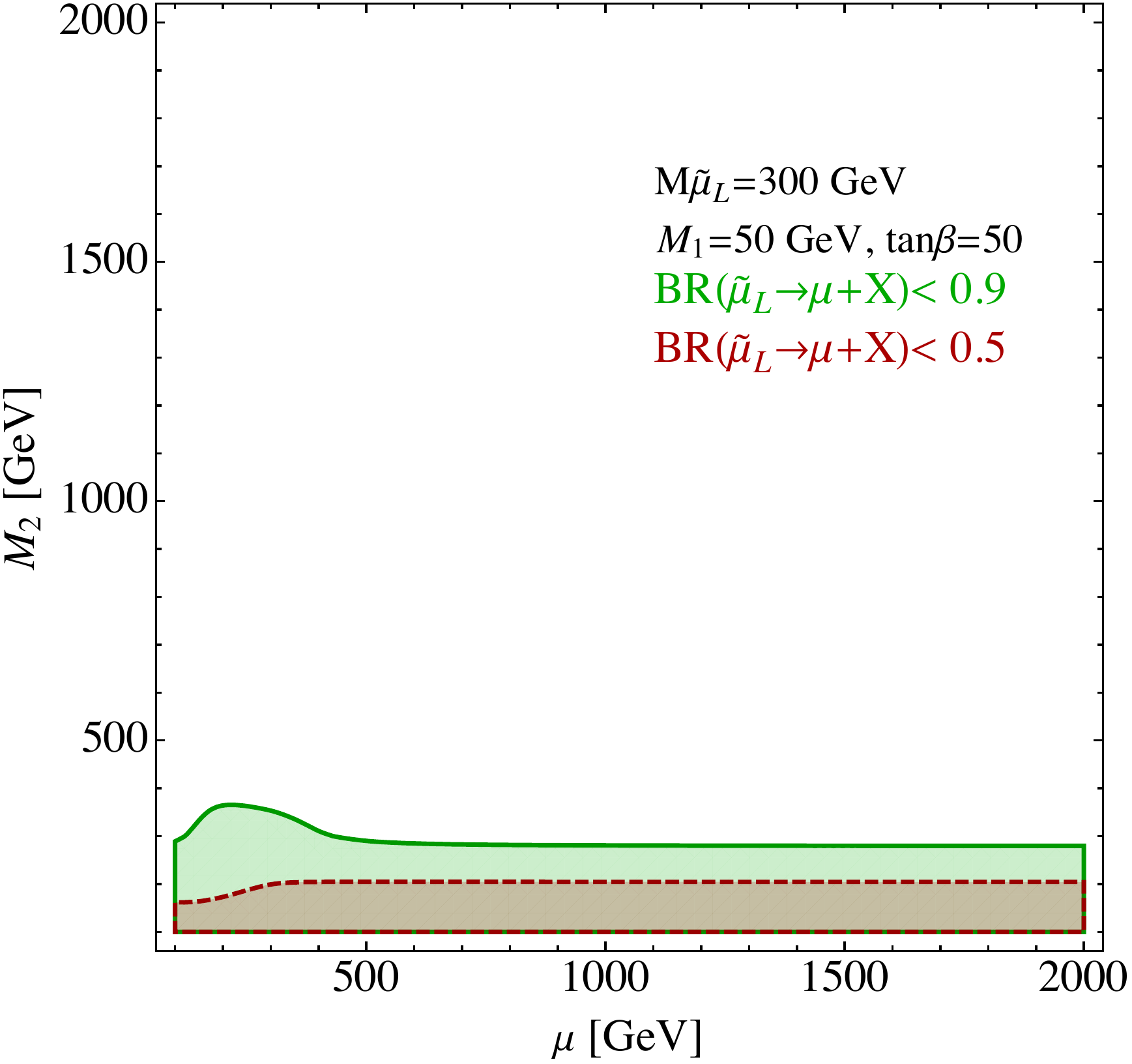}
\caption{Parameter space in the $\mu-M_2$ plane where ${\rm BR}({\tilde{\mu}^\pm_{L}} \rightarrow \mu^\pm + X)$ deviates from one, for $M_{\smuL}=300 \gev$. The green solid (red dashed) area encloses the region where  ${\rm BR}({\tilde{\mu}^\pm_{L}} \rightarrow \mu^\pm + X)<0.9 ~(0.5)$. In the region above the green solid line (the white space)  ${\rm BR}({\tilde{\mu}^\pm_{L}} \rightarrow \mu^\pm + X)$ is greater than 0.9.}
\label{fig:smuon_BR}
\end{figure}

As long as the $\C1$ and $\N2$ are heavier than $\smuL$, the above assumption of ${\rm BR}({\tilde{\mu}^\pm_{L}} \rightarrow \mu^\pm \neu) = 1$ remains valid. However, in the MSSM parameter space of our interest, there are parameter regions where  $\C1$ and $\N2$ can be lighter than $\smuL^\pm$, and it is important to determine to what extent the above lower bound of $300\gev$ can be modified. Taking for example a scenario with $M_1=50\gev$ and $\tan \beta=50$, we show by the green (red) shaded area in Fig.~\ref{fig:smuon_BR} the parameter region in the $\mu-M_2$ plane in which the ${\rm BR}({\tilde{\mu}^\pm_{L}} \rightarrow \mu^\pm + X) < 0.9 ~(0.5)$, where $X$ stands for anything other than the $\mu^\pm$ originating from the decay of $\smuL^{\pm}$. We find only a very small region of parameter space where the ${\rm BR}({\tilde{\mu}^\pm_{L}} \rightarrow \mu^\pm + X)$ is smaller than $50\%$. The smuon decay BR to muons here includes contributions from the processes $\smuL^\pm \rightarrow \mu^\pm + \neu$, as well as from $\smuL^\pm \rightarrow \mu^\pm + \N2$ (and subdominantly from $\smuL^{\pm} \rightarrow \nu_{\mu} + \C1$, where subsequently $\C1 \rightarrow W^\pm (\rightarrow \mu^\pm \nu_\mu) \neu$). As we can see from Fig.~\ref{fig:smuon_BR}, the branching to $\C1$ is enhanced only when the $\C1$ is wino-like and light. In order to obtain the same event yield that can be excluded by the smuon search under consideration, the reduction in BR to final states containing muons can be compensated by an enhanced cross-section for smuon pair production. We can thus translate the $95\%$ C.L. lower bound on the left-smuon mass, 300 GeV in Eq.~\ref{smuon_bound}, as follows:
\begin{align}
M_{\smuL} &> 287 \gev, \text{~~~for~} {\rm BR}({\tilde{\mu}^\pm_{L}} \rightarrow \mu^\pm + X) = 0.9, \nonumber \\
M_{\smuL} &> 220 \gev, \text{~~~for~} {\rm BR}({\tilde{\mu}^\pm_{L}} \rightarrow \mu^\pm + X) = 0.5~.
\end{align}

With the above lower bound on the left-scalar muon mass, the requirement of explaining the $\gm2$ anomaly by the chargino contribution would imply an upper bound on the chargino mass. The goal of the subsequent sections is thus to determine whether the current LHC searches allow for a chargino lighter than that upper bound. As mentioned in Sec.~\ref{sec:gm2}, the MSSM contribution to $\gm2$ from diagrams involving the right-smuon is not discussed in this study, and we therefore take the right-smuon to be decoupled from the spectrum.

\section{Chargino mass limits from the LHC}
\label{sec:chargino}
As discussed in the Introduction, in a completely general MSSM scenario, all the electroweak sector mass and mixing parameters are relevant for interpreting the different LHC searches for the electroweak-inos. Thus, performing a global analysis of the MSSM parameter space, including several different LHC search channels, requires very large computational resources, and is beyond the scope of the present study. However, as we shall demonstrate in the following, to draw certain broad conclusions on the $\gm2-$compatible parameter space allowed by current constraints, performing a smaller subset of analyses is sufficient. 

In the interpretation of different searches for $\tilde{\chi}_1^+ \tilde{\chi}_1^-$ and $\C1 \N2$ production carried out by the LHC collaborations, the hierarchies between the $\C1/\N2$ and slepton masses play a crucial role. Assuming for simplicity $M_{\smuL}=M_{\tilde{e}_L}$, and for fixed values of $M_{\tilde{\tau}_2}$ and  $\theta_{\tilde{\tau}}$ (which, as we shall see later, are less relevant parameters), we then have the following possible mass hierarchies for a bino-like LSP ($\neu$) scenario:
\begin{enumerate}
\item $M_{\smuL}>M_{\C1},M_{\N2}>M_{\tilde{\tau}_1}$

\item $M_{\tilde{\tau}_1}>M_{\C1},M_{\N2}>M_{\smuL}$

\item $M_{\C1},M_{\N2}>M_{\tilde{\tau}_1},M_{\smuL}$ (with either hierarchy between $\tla$ and $\smuL$) 

\item $M_{\smuL},M_{\tilde{\tau}_1}>M_{\C1},M_{\N2}$.
\end{enumerate}
In the following subsections, we take up each of the above hierarchies in turn, and discuss the constraints on the MSSM parameters of interest from the relevant LHC searches.

\subsection{Scenario-1: stau NLSP}
\label{stau}
We first focus on the mass hierarchy $M_{\smuL}>M_{\C1},M_{\N2}>M_{\tilde{\tau}_1}>M_{\neu}$. In this scenario, the $\C1$ and $\N2$ would decay dominantly via the intermediate $\tla$, the tau-sneutrino, and $\tlb$ (when kinematically accessible), and subdominantly through $W/Z/h$, thereby leading to a multi-tau final state. Considering hadronic decays of the tau leptons ($\tau_h$), the final state of interest is then $\geq 2\tau_h+\met$. We follow the corresponding ATLAS search strategies~\cite{ATLAS_tau_8, ATLAS_tau_13} in this regard, on which our constraints are based. Utilizing both the  $20.3 \fb^{-1}$ data from the 8 TeV LHC run~\cite{ATLAS_tau_8} and the $14.8 \fb^{-1}$ of data from 13 TeV~\cite{ATLAS_tau_13}, the ATLAS collaboration has looked for $\C1 \N2$ production, and interpreted the results within a simplified model setup, assuming $M_{\tla,\tilde{\nu_\tau}}=(M_{\neu}+M_{\N2})/2$. The LHC8 and LHC13 searches lead to the following limit from ATLAS:
\begin{equation}
M_{\C1} = M_{\N2} > 700 \gev, \text{~for~} M_{\neu}=0 \gev, \text{~at~}95\% \text{~C.L.}
\end{equation}

Interpreting the search with at least two hadronically decaying tau leptons and $\met$ is considerably involved, as the crucial tau jet identification and reconstruction efficiencies are sensitive to the kinematics of the tau jets, and thus in turn to the mass splittings between the sparticles. We therefore study in detail how the LHC bounds translate to constraints in the corresponding MSSM parameter space, taking into account both modifications to the relevant production rates and BRs, as well as by performing a Monte Carlo study with a simple detector simulation to capture the possible changes in the detection efficiencies. The details of our MC simulation framework, the kinematic selection criteria employed, as well as the validation of our simulation framework are discussed in the Appendix. We note here that due to our modelling of the tau reconstruction and identification efficiencies based on a simple detector simulation, our event yields are in general larger than that reported by ATLAS. We therefore rescale our event yields to match the ATLAS numbers by a constant fudge factor each for the 8 and 13 TeV searches, as detailed in the Appendix. 

In our analysis, we have incorporated certain simple but important effects not captured by the above simplified model study by ATLAS. To begin with, we include all possible electroweak-ino production modes, namely, 
\begin{align}
 p p &\rightarrow \tilde{\chi}_i^+\tilde{\chi}_j^-, \text{~with~} i, j =1,2,\nonumber \\
       &\rightarrow \tilde{\chi}_i^\pm\tilde{\chi}_j^0, \text{~with~} i =1,2, \text{~and~} j=1,2,3,4,\nonumber \\
       &\rightarrow \tilde{\tau}_k^+\tilde{\tau}_\ell^-, \text{~with~} k,\ell=1,2. 
\end{align}        
Clearly, not all of the production processes have appreciable cross-sections for a given set of parameters, but there are regions of parameter space (for example, for Higgsino-like $\tilde{\chi}_2^0$ and $\tilde{\chi}_3^0$), where including the production of heavier charginos and neutralinos is important. We do not make any assumptions about the elctroweak-ino decay branchings, and take into account their decays via staus, tau-sneutrino as well as via the SM gauge and Higgs bosons, according to the relevant BRs. Finally, the impact of variation in the stau mixing, the LSP mass, as well as $\tan \beta$ are also studied. 

\begin{figure}[htb!]
\centering
\includegraphics[scale=0.6]{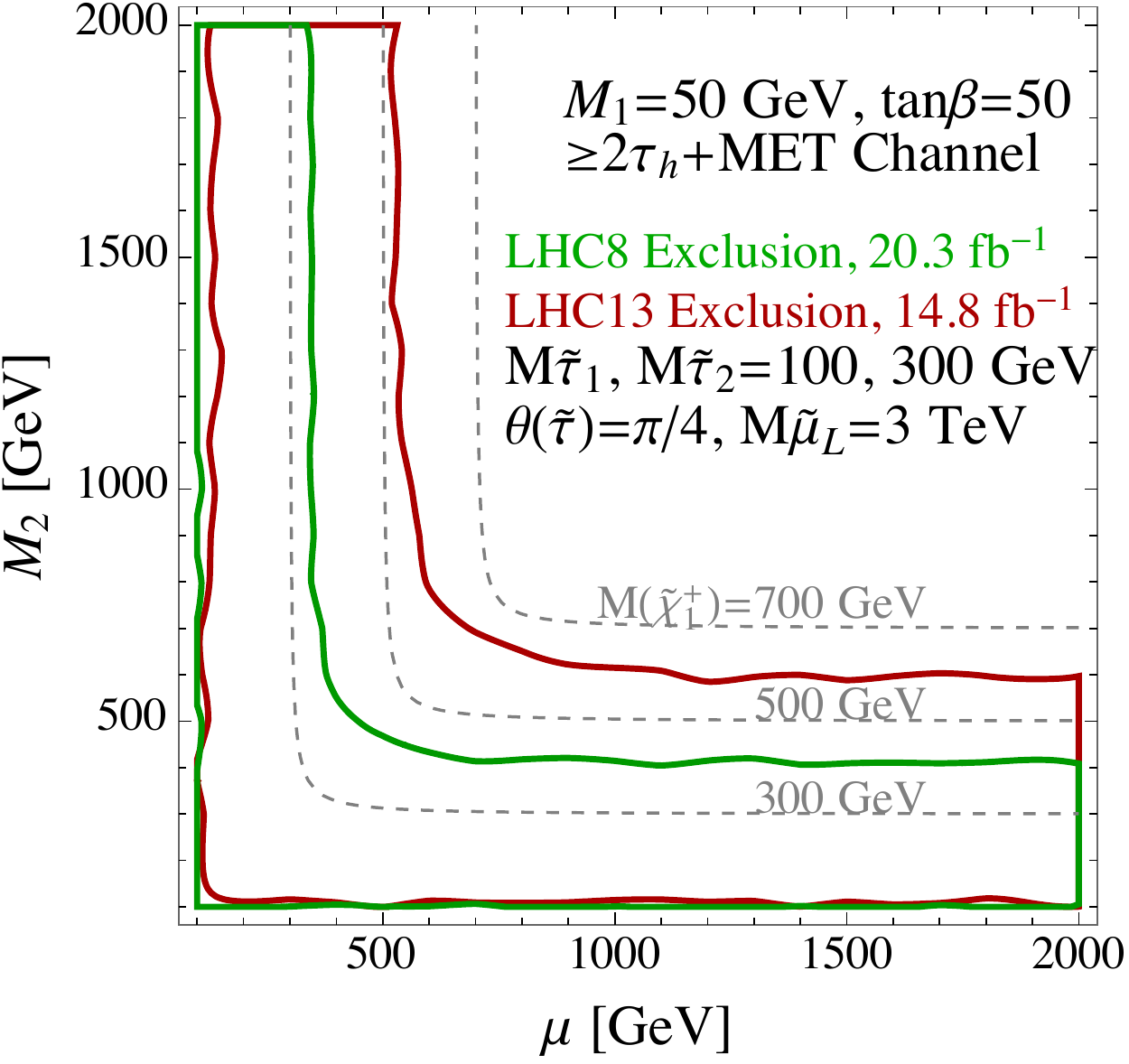}
\caption{\small\sl $95\%$ C.L. exclusion in the $\mu-M_2$ plane following the ATLAS search in the $\geq 2\tau_h+\met$ channel, using the 8 TeV, $20.3 \fb^{-1}$ data (green) and 13 TeV, $14.8 \fb^{-1}$ data (red) from the LHC. The results are shown for a scenario where the left-smuon is decoupled ($M_{\smuL}=3$ TeV), with $M_1=50 \gev, \tan \beta=50, M_{\tla}=100 \gev, M_{\tlb}=300 \gev$ and $\theta_{\tilde{\tau}}=\pi/4$. We set $M_{\tilde{e}_L}=M_{\smuL}$ for simplicity, and the right selectron and right smuon, which are not relevant for this study, are taken to be decoupled.}
\label{fig:tau_search}
\end{figure}
For the dominant chargino contribution to the $\gm2$, apart from the left-smuon mass and $\tan\beta$, the two other important parameters are $\mu$ and $M_2$, which also determine the chargino and heavier neutralino production and decay rates. Therefore, we demonstrate the LHC constraints in the $\mu-M_2$ parameter plane. In Fig.~\ref{fig:tau_search}, we show the $95\%$ C.L. excluded region in this plane using the 8 TeV, $20.3 \fb^{-1}$ data (green) and 13 TeV, $20.3 \fb^{-1}$ data (red) from the LHC following the ATLAS search in the $\geq 2\tau_h+\met$ channel. The results are shown for the left-smuon decoupled scenario ($M_{\smuL}=3$ TeV), with $M_1=50 \gev, \tan \beta=50, M_{\tla}=100 \gev, M_{\tlb}=300 \gev$ and $\theta_{\tilde{\tau}}=\pi/4$. While the 8 TeV search covers $\C1$ masses of upto around 400 GeV, this reach is significantly extended with the currently analyzed 13 TeV dataset, namely $\C1$ masses of upto around 600 GeV are now excluded across the $\mu-M_2$ plane. Since for the same mass values, the wino-like $\C1 \N2$ production has a higher cross-section than the Higgsino-like one (by a factor of four), the exclusion is slightly stronger for high $\mu$ and smaller $M_2$ values. The cross-section reduction in the Higgsino-like region is, however, partially compensated by the associated production of additional Higgsino-like heavier neutralino states of similar mass with a $\C1$. We find that apart from the very light $\C1/\N2$ mass regions with low $\mu$ and high $M_2$ values, the 13 TeV search also covers the region excluded by the 8 TeV analysis. With this in mind, in subsequent figures, we shall only show the 13 TeV exclusion contours. 

\begin{figure}[htb!]
\centering
\includegraphics[scale=0.6]{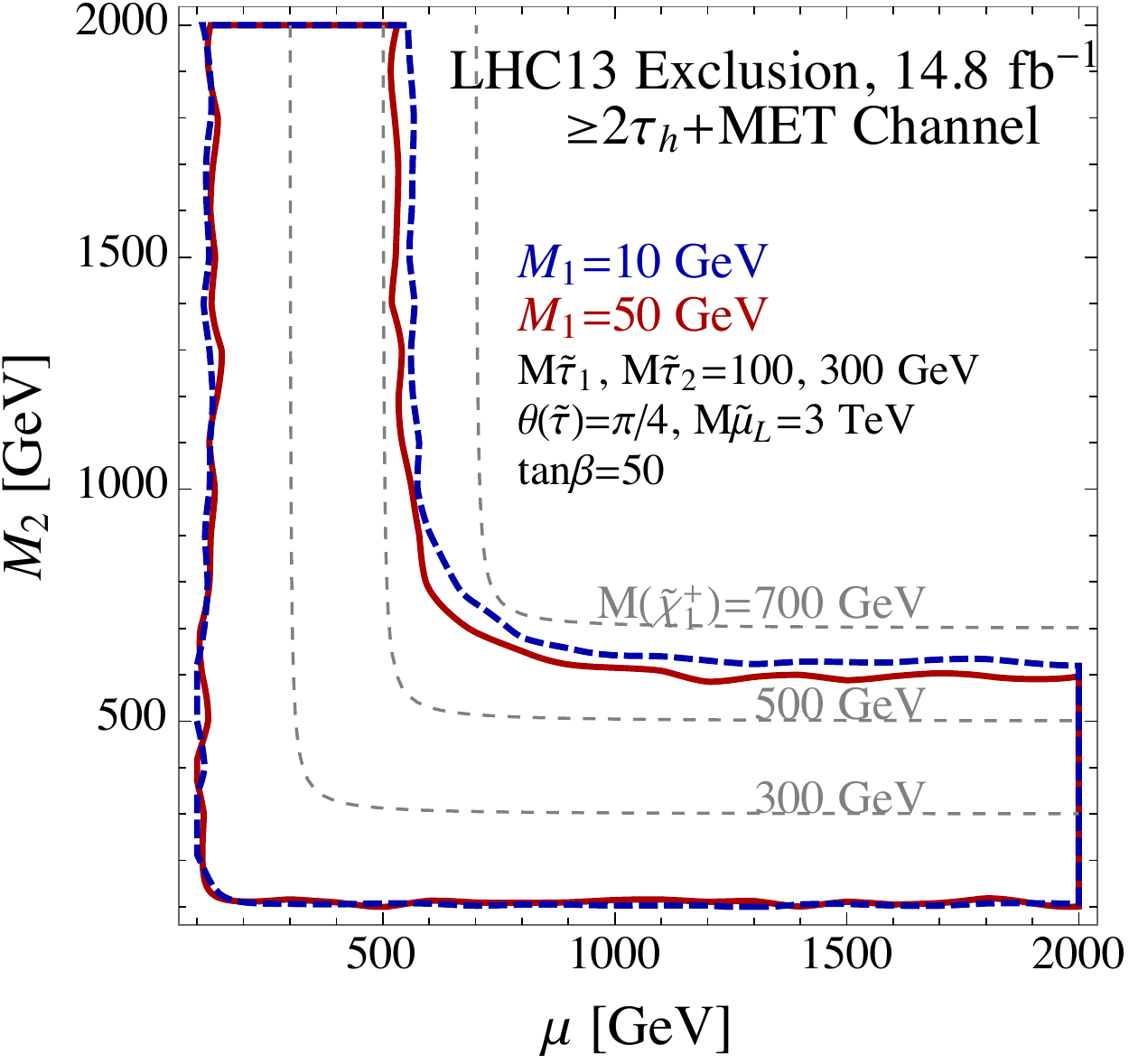}
\caption{\small\sl $95\%$ C.L. exclusion in the $\mu-M_2$ plane for different values of the LSP mass, $M_1=10\gev$ (blue-dashed line) and $M_1=50 \gev$ (red solid line). The exclusion contours are based on the 13 TeV, $14.8 \fb^{-1}$ data from the LHC, following the ATLAS search in the $\geq 2\tau_h+\met$ channel. The results are shown for a scenario where the left-smuon is decoupled ($M_{\smuL}=3$ TeV), with $\tan \beta=50, M_{\tla}=100 \gev, M_{\tlb}=300 \gev$ and $\theta_{\tilde{\tau}}=\pi/4$. We set $M_{\tilde{e}_L}=M_{\smuL}$ for simplicity, and the right selectron and right smuon, which are not relevant for this study, are taken to be decoupled.}
\label{fig:M1_vary}
\end{figure}
In order to demonstrate the impact of change in kinematic cut efficiencies due to different values of the $\neu$ LSP mass, we show in Fig.~\ref{fig:M1_vary} the comparison of the current LHC 13 TeV exclusions for $M_1=10 \gev$ (blue dashed) and $M_1=50 \gev$ (red solid). The rest of the parameters are fixed as in Fig.~\ref{fig:tau_search}. As we can see from this figure, the impact of varying $M_1$ is minimal, and as long as the bino-like $\neu$ is not degenerate with the $\tla$ NLSP, the kinematic cut efficiencies do not vary significantly. Therefore, in what follows, we shall show all of our results for the $M_1=50\gev$ scenario. 

\begin{figure}[htb!]
\centering
\includegraphics[scale=0.6]{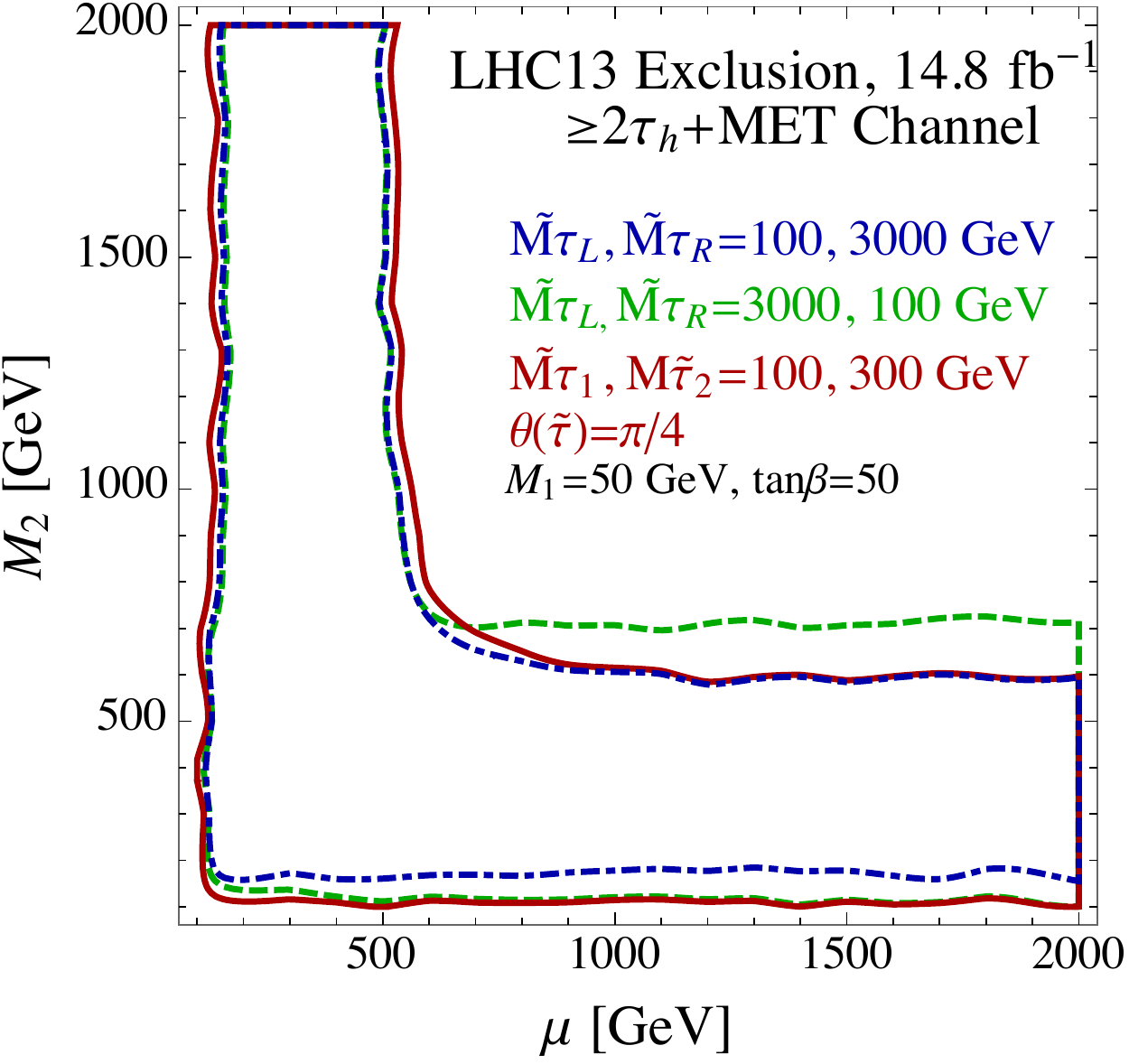}
\caption{\small\sl $95\%$ C.L. LHC13 exclusion contours for different values of the mixing angle in the stau sector. The red (solid), green (dashed) and blue (dot-dashed) lines correspond to the lighter stau being maximally mixed, right-stau and left-stau respectively. The other relevant parameters are fixed as in Fig.~\ref{fig:tau_search}.}
\label{fig:stau_mix}
\end{figure}
We now consider the impact of varying the mixing angle in the stau sector on the LHC exclusions. In Fig.~\ref{fig:stau_mix}, we show results for three different choices for $\theta_{\tilde{\tau}}$: the lighter stau being a left-stau ($\theta_{\tilde{\tau}}=0$), a right-stau ($\theta_{\tilde{\tau}}=\pi/2$) or an equal admixture of both ($\theta_{\tilde{\tau}}=\pi/4$). In the first (second) scenario, the right (left) stau mass is set to be $3\tev$. For the first two cases, though we set the trilinear parameter $A_\tau = 0$, there is always a small mixing between the left and right stau due to non-zero values of $\mu \tan \beta$. When the lighter stau is a $\tilde{\tau}_L$, the tau sneutrino is also light and of similar mass, and therefore decays through the tau sneutrino leads to a reduction in ${\rm BR}(\N2 \rightarrow \tau^+ \tau^- \neu)$. This reduction in the BR to tau-rich final states is also applicable to the maximally mixed scenario, as decays through the sneutrino state are still accessible, albeit with a slightly smaller branching (due to a somewhat higher mass of the sneutrino compared to the $\tla$). When the lighter stau is a right-stau with the left-stau mass set at 3 TeV, the sneutrino is not kinematically accessible, thereby enhancing the BR of $\N2$ to tau's, and hence increasing the exclusion reach in the wino-like region ($M_2<\mu$). In the Higgsino like region ($\mu<M_2$) there is no difference observed, as a Higgsino-like $\N2$ does not decay via sneutrinos. Keeping in mind that the $\theta_{\tilde{\tau}}=45^\circ$ scenario leads to slightly weaker bounds compared to the right-stau NLSP case, for subsequent figures we shall show results for the maximally mixed scenario only. 

We note in passing that a left-right mixed $\tla$ NLSP with a bino-like LSP is well-motivated from the point of view of obtaining a DM candidate with the required thermal relic abundance (for $M_1 \geq \mathcal{O}(30 \gev)$), or for accommodating the Galactic Centre gamma ray excess through $\neu \neu \rightarrow \tau^+ \tau^-$ annihilation in the present epoch (for $M_1 \sim 10 \gev$)~\cite{SM}.

\subsection{Scenario-2: selectron/smuon NLSP}
\label{trilepton}
The second mass hierarchy we consider is $M_{\tilde{\tau}_1}>M_{\C1},M_{\N2}>M_{\smuL}, M_{\tilde{e}_L}>M_{\neu}$. In this case, the $\C1$ and $\N2$ would decay dominantly through the $\smuL$ and $\tilde{e}_L$, their corresponding sneutrinos, and subdominantly through $W/Z/h$. Assuming a $100\%$ decay through $\smuL,\tilde{e}_L$ and their associated sneutrinos, the current ATLAS bounds from $\C1 \N2$ search in the $3\ell+\met$ channel (with $\ell=e^\pm,\mu^\pm$) using $13.3 \fb^{-1}$ of data from the 13 TeV LHC is as follows~\cite{ATLAS_C1N2_13}:
\begin{equation}
M_{\C1} = M_{\N2} > 1000 \gev, \text{~for~} M_{\neu}=0 \gev, \text{~at~}95\% \text{~C.L.}
\end{equation}
For this ATLAS analysis, the $\C1$ and $\N2$ are assumed to be wino-like, while the $\neu$ is taken to be bino-like. These limits are expected to become slightly weaker in the Higgsino-like or wino-Higgsino mixed region for $\C1/\N2$.
\begin{figure}[htb!]
\centering
\includegraphics[scale=0.6]{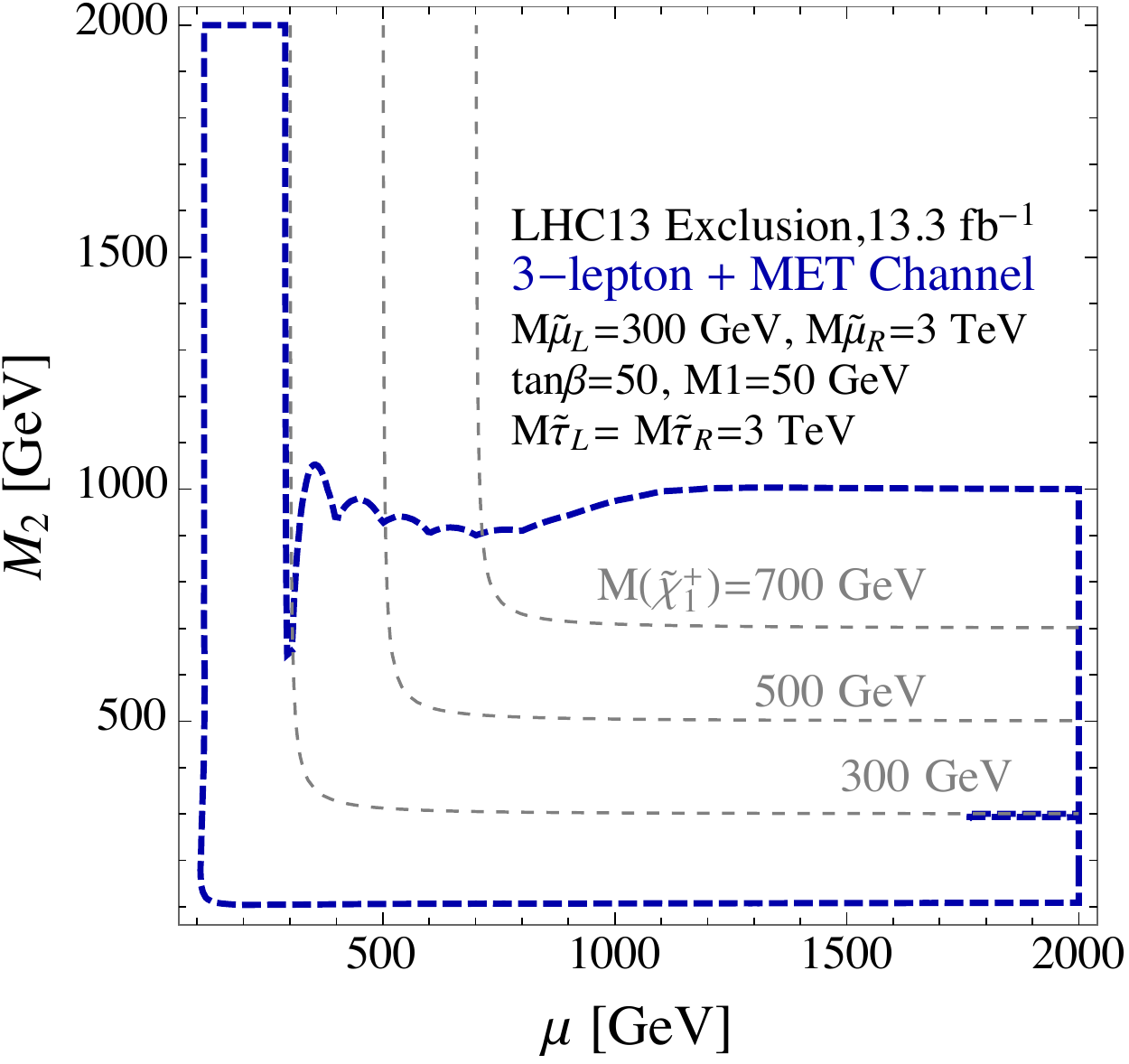}
\caption{\small\sl Estimated $95\%$ C.L. exclusion region in the $\mu-M_2$ plane using the 13 TeV, $13.3 \fb^{-1}$ data (blue dashed line) from the LHC, following the ATLAS search in the $3\ell+\met$ channel. The results are shown for the scenario with both the staus decoupled ($M_{\tilde{\tau}_L}=M_{\tilde{\tau}_R}=3$ TeV), with $M_1=50 \gev, \tan \beta=50$ and $M_{\smuL}=300 \gev$. We set $M_{\tilde{e}_L}=M_{\smuL}$ for simplicity, and the right selectron and right smuon, which are not relevant for this study, are taken to be decoupled.}
\label{fig:trilepton_all}
\end{figure}

Unlike in the stau-NLSP scenario discussed in the previous sub-section, we do not perform a detailed MC analysis of the kinematic selection efficiencies and detector effects for this mass hierarchy. This should not lead to significant modifications to the derived constraints, since in the very clean trilepton and $\met$ channel, the final state reconstruction and detector resolution effects are expected to be small, as long as we do not encounter degeneracies between the electroweak-ino, slepton and bino mass parameters. For example, though the above lower bound is quoted for $M_{\neu}=0 \gev$, it remains essentially unchanged in the range $0 \gev \leq M_{\neu} \lesssim 500 \gev$~\cite{ATLAS_C1N2_13}. Similarly, though the smuon/selectron mass for the simplified model based search is assumed to be given by $M_{\smuL,\tilde{e}_L}=(M_{\neu}+M_{\N2})/2$, it has been shown in Ref.~\cite{ATLAS_EW_Summary} that away from the degenerate mass region, this mass gap does not affect the limits significantly. Therefore, for our discussion in this sub-section, we shall only consider the production cross-section and decay BR modifications of $\C1$ and $\N2$ in the $\mu-M_2$ plane.

In Fig.~\ref{fig:trilepton_all}, we show our estimate of the current 13 TeV exclusions in the $\mu-M_2$ plane using the trilepton channel (blue dashed curve), for $M_{\smuL,\tilde{e}_L}=300 \gev$, with the $\tilde{\tau}_L, \tilde{\tau}_R, \tilde{e}_R$ and $\tilde{\mu}_R$ decoupled (their masses are fixed at 3 TeV). The results are shown for $M_1=50\gev$ and $\tan\beta=50$. The smuon mass is set at its lower bound from 8 TeV LHC searches, as discussed in Sec.~\ref{smuon}, which is valid in almost the entire parameter space except for very small $M_2$ values, where it can be slightly lower. 

We observe two branches in the blue-dashed exclusion contour in Fig.~\ref{fig:trilepton_all}. For the region where the $\C1/\N2$ is lighter than $\smuL/\tilde{e}_L$, they decay via the SM gauge and Higgs bosons, thereby reducing the BR to the trilepton final state. Therefore, only smaller $\C1/\N2$ mass values are accessible to the trilepton search when only the gauge/Higgs boson decay modes are open. For $M_{\C1}(=M_{\N2})$ larger than 300 GeV, decays through intermediate sleptons open up, and higher electroweak-ino masses can be probed as well. However, for the latter case, the bounds are weaker for Higgsino-like $\C1/\N2$ region compared to the wino-like one, as only the wino component of the $\C1/\N2$ has significant coupling to the left-smuon or the muon-sneutrino.

\subsection{Scenario-3: all three generation sleptons lighter than chargino}
\label{sec:sc3}
\begin{figure}[htb!]
\centering
\includegraphics[scale=0.6]{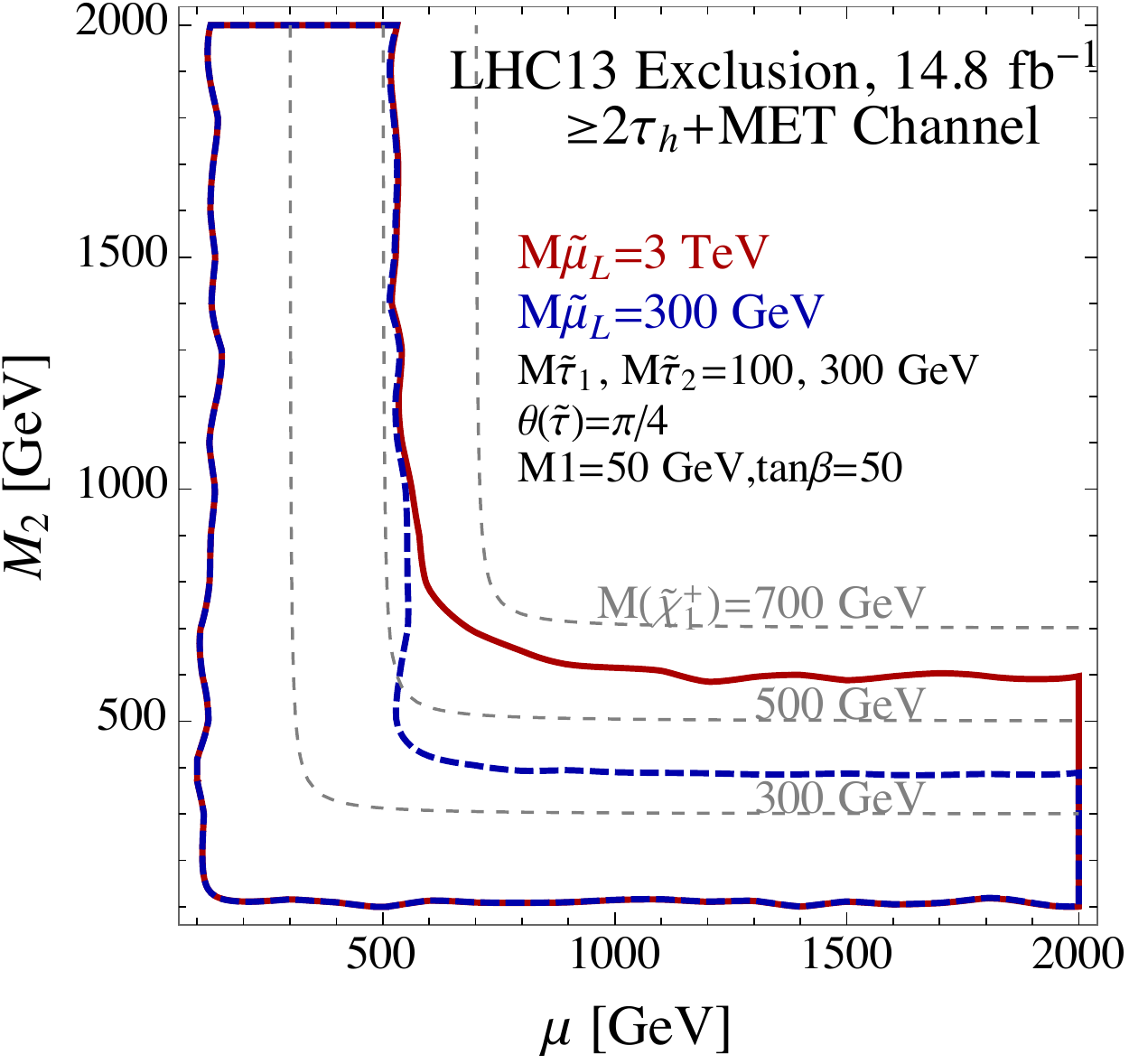}
\caption{\small\sl $95\%$ C.L. LHC13 exclusion contours, using the $\geq 2\tau_h+\met$ search only, for different $\smuL$ mass values. The red (solid) and blue (dashed) lines correspond to $M_{\smuL}=3000 \gev$ and $M_{\smuL}=300 \gev$ respectively. We set $M_{\tilde{e}_L}=M_{\smuL}$ for simplicity.}
\label{fig:smuon_mass}
\end{figure}

We next study the mass hierarchy: $M_{\C1},M_{\N2}>M_{\tilde{\tau}_1},M_{\smuL,\tilde{e}_L}>M_{\neu}$ (with either hierarchy between $\tla$ and $\smuL$), which presents us with a scenario where both the $\geq 2\tau_h+\met$ and the $3\ell+\met$ searches described in the previous two sub-sections become important. Let us first consider how the bounds shown in Fig.~\ref{fig:tau_search} get modified in the presence of a light left-smuon kinematically accessible in $\C1/\N2$ decays. In Fig.~\ref{fig:smuon_mass}, we show the LHC13 exclusion contours, using the $\geq 2\tau_h+\met$ search only, for two different $\smuL$ mass values. The red (solid) and blue (dashed) lines correspond to $M_{\smuL}=3000 \gev$ and $M_{\smuL}=300 \gev$ respectively, while the other relevant parameters are the same as in Fig.~\ref{fig:tau_search}. It is observed that for the wino-like $\C1$ and $\N2$ region, there is a decrease in the reach of the tau search channel, since the wino-like $\C1/\N2$ will also have a significant decay branching to smuon and muon sneutrino states, thereby reducing the BR to the multi-tau final state. 

\begin{figure}[htb!]
\centering
\includegraphics[scale=0.6]{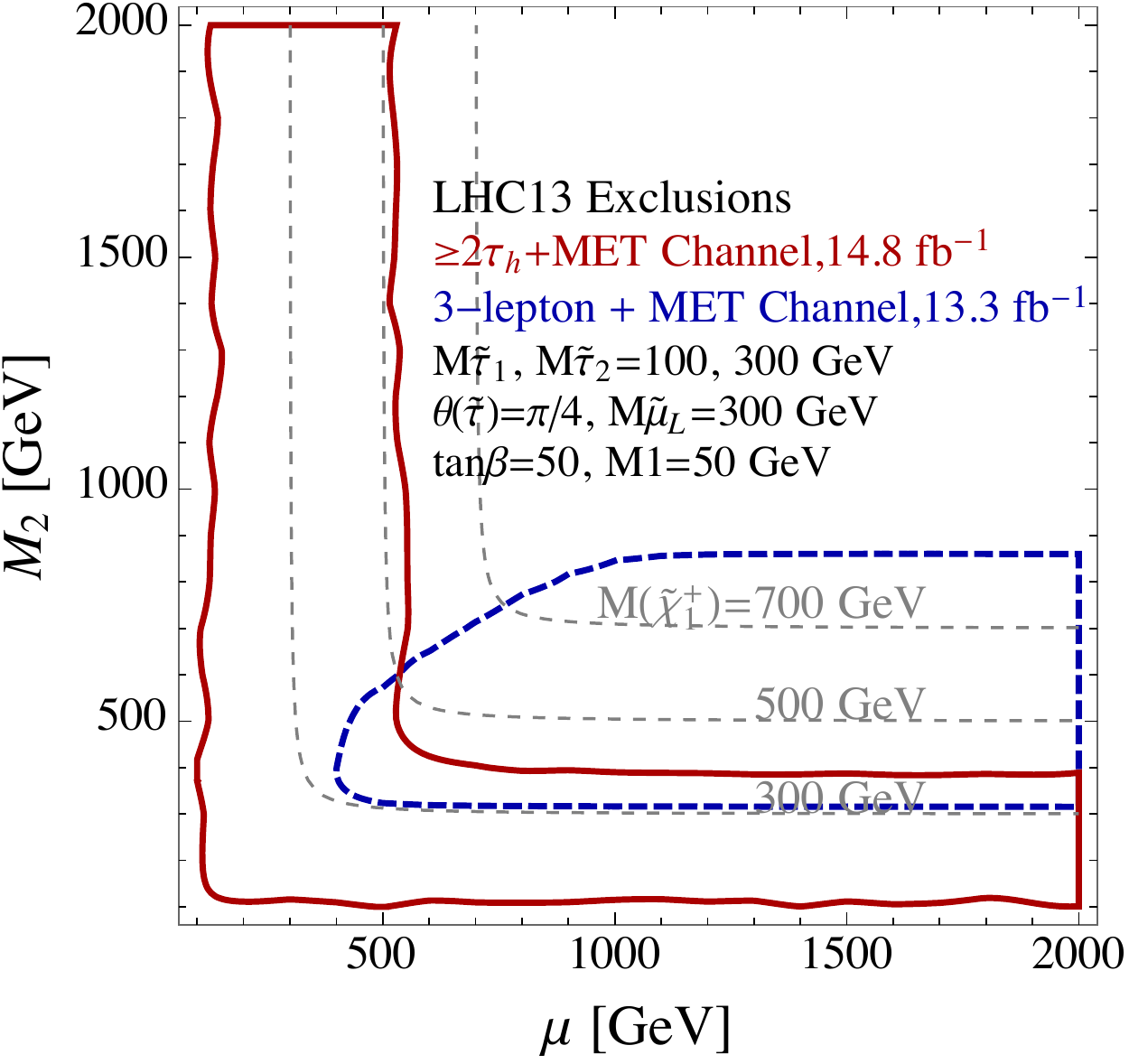}
\caption{\small\sl $95\%$ C.L. LHC13 exclusion contours, using the $\geq 2\tau_h+\met$ search (red solid) and the $3\ell+\met$ search (blue dashed), in a scenario where both $\tla$ and $\smuL$ can be lighter than the $\C1/\N2$. The results are shown for $\tan \beta=50, M_1=50 \gev, M_{\smuL}=300 \gev, M_{\tla}=100 \gev, M_{\tlb}=300 \gev$ and $\theta_{\tilde{\tau}}=\pi/4$. We set $M_{\tilde{e}_L}=M_{\smuL}$ for simplicity. For most regions of the parameter space, the two searches are found to be complementary.}
\label{fig:stau_smuon}
\end{figure}

The region in which the $\geq 2\tau_h+\met$ search loses power, the $3\ell+\met$ search becomes sensitive. In general, a statistical combination of both these search channels should lead to the strongest constraints in the $\mu-M_2$ plane. There is, however, a degree of complementarity to the power of these two searches. We demonstrate this fact in Fig.~\ref{fig:stau_smuon}, which shows that we can separate different areas in the $\mu-M_2$ plane in which one of the two searches has a higher exclusion power. For reasons discussed above, in the lower $\mu$ and higher $M_2$ region, the multi-tau channel is stronger, while in the lower $M_2$ and higher $\mu$ region, the multilepton channel dominates. Along the diagonal, with $\mu \sim M_2$, both searches become less powerful, especially for $M_{\C1}>500 \gev$. 

\subsection{Scenario-4: all sleptons heavier than chargino}
\label{no_slep}
Finally, we comment on the mass hierarchy in which $M_{\smuL},M_{\tilde{\tau}_1}>M_{\C1},M_{\N2}>M_{\neu}$. Since in this case all the sleptons are heavier than $\C1$ and $\N2$, the latter can only decay via $W/Z/h$ bosons. Based on $20.3 \fb^{-1}$ of data from the 8 TeV LHC run (the 13 TeV analysis for these $\C1/\N2$ decay modes are not yet published by the LHC collaborations), the ATLAS collaboration obtains the following limit using a combination of dilepton and trilepton searches~\cite{ATLAS_Slepton}:
\begin{equation}
M_{\C1} = M_{\N2} > 415 \gev, \text{~for~} M_{\neu}=0 \gev, \text{~at~}95\% \text{~C.L.}~~\text{(WZ~mode)}.
\end{equation}
In deriving the above limit, it is assumed that $\text{BR}(\C1 \rightarrow W^\pm \neu)=1$ and $\text{BR}(\N2 \rightarrow Z \neu)=1$. If on the other hand, one adopts a simplified model in which $\text{BR}(\C1 \rightarrow W^\pm \neu)=1$ and $\text{BR}(\N2 \rightarrow h \neu)=1$, the above bound becomes considerably weaker, and a combination of search results targeting different Higgs decay modes yields the following bound from ATLAS (using $20.3 \fb^{-1}$ of data from the 8 TeV LHC run)~\cite{ATLAS_Wh}:
\begin{equation}
M_{\C1} = M_{\N2} > 250 \gev, \text{~for~} M_{\neu}=0 \gev, \text{~at~}95\% \text{~C.L.}~~\text{(Wh~mode)}.
\end{equation}
The assumption of either $\text{BR}(\N2 \rightarrow Z \neu)=1$ or $\text{BR}(\N2 \rightarrow h \neu)=1$ does not hold in most regions of the $\mu-M_2$ parameter space, and a combination of $\N2 \rightarrow Z \neu$ and $\N2 \rightarrow h \neu$ decays take place depending upon the Higgsino component of $\N2$ and $\neu$. We do not study this mass hierarchy in this paper, and ideally a statistical combination of the searches targeting the $Z\neu$ and $h\neu$ decay modes of $\N2$ should be performed in order to determine the current best constraints in the $\mu-M_2$ plane for this scenario. We note that although this scenario leads to the weakest bounds on the electroweak-inos, making the sleptons, especially the left-smuon heavier, is not favourable to the chargino contribution to the $\gm2$. However, it is expected that a part of the $\gm2$ favoured parameter space would be allowed for this mass hierarchy, even after the 8 TeV LHC constraints are taken into account.

\section{Impact of LHC constraints on the ${\gm2}$ favoured parameter space}
\label{combi}
We are now in a position to discuss the impact of the LHC constraints on the MSSM parameter space in which the $\gm2$ anomaly can be explained within $1\sigma$. As emphasized in Sec.~\ref{sec:gm2}, we focus on the right-smuon decoupled scenario, where the dominant contribution to $\Delta a_\mu$ comes from the chargino-muon sneutrino loop. For our numerical analysis, we have included all the relevant chargino and neutralino contributions to $\Delta a_\mu$, with $M_{\tilde{\mu}_R}$ fixed at 3 TeV.
\subsection{Scenario-1: stau NLSP}
\begin{figure}[htb!]
\centering
\includegraphics[scale=0.6]{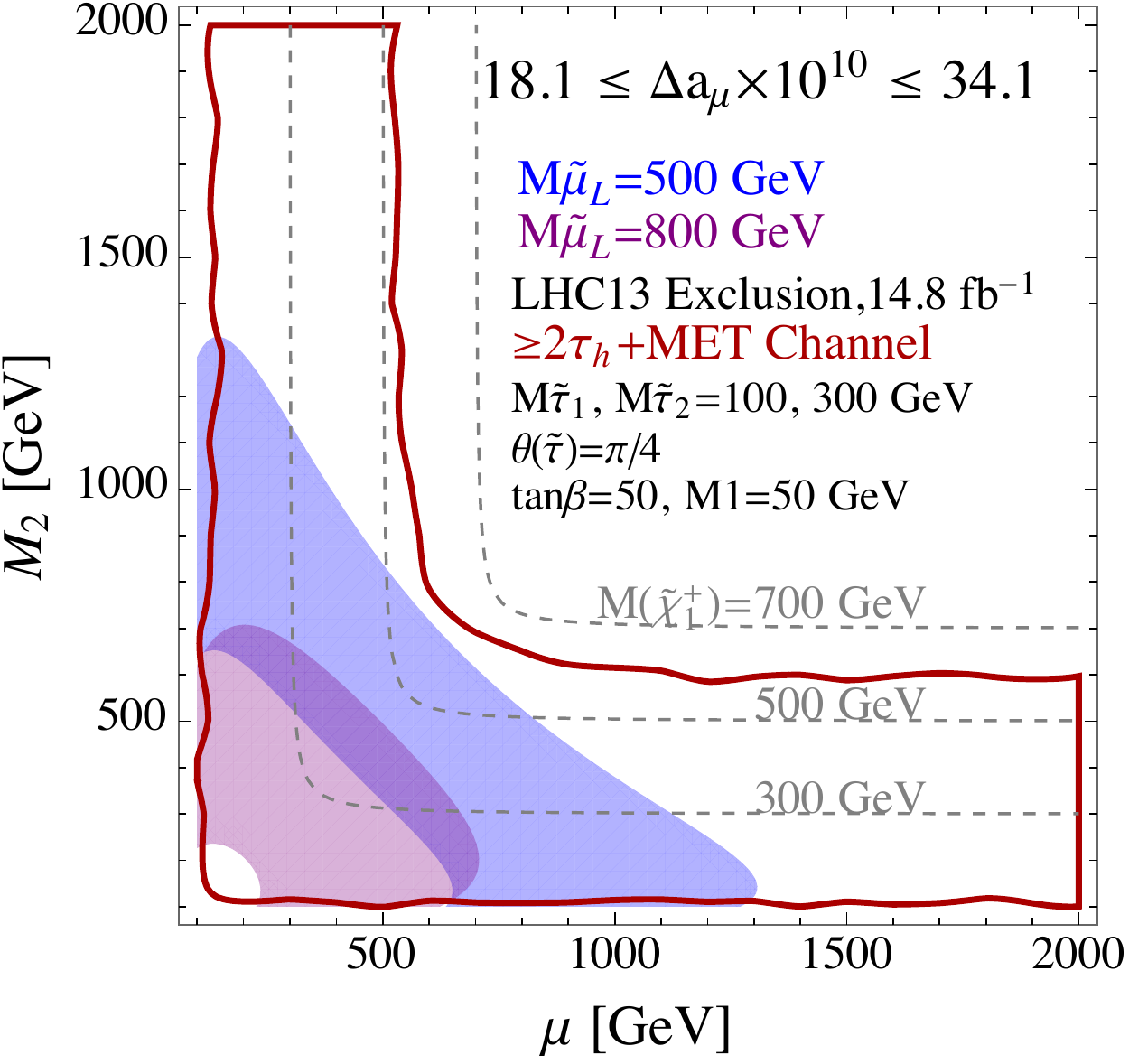}
\caption{\small\sl Current $95\%$ C.L. LHC constraints on the $\gm2$ favoured parameter region in the $\mu-M_2$ plane, for the mass hierarchy $M_{\smuL},M_{\tilde{\tau}_1}>M_{\C1},M_{\N2}>M_{\neu}$. In the blue (violet) shaded region, the $\gm2$ anomaly can be explained at $1\sigma$, with the choice $M_{\smuL}=500 \gev ~(800 \gev)$. The values of other relevant MSSM parameters are fixed as in Fig.~\ref{fig:tau_search}.}
\label{fig:gm21}
\end{figure}

We start with the first mass hierarchy : $M_{\smuL}>M_{\C1},M_{\N2}>M_{\tilde{\tau}_1}>M_{\neu}$. In Fig.~\ref{fig:gm21}, we show two representative scenarios for the $\gm2$ favoured region at $1\sigma$, with $M_{\smuL}=500 \gev$ (blue shaded) and 800 GeV (violet shaded). As we can see in Fig.~\ref{fig:gm21}, for $M_{\smuL}=500 \gev$, the above mass hierarchy assumption is not valid in a small region of the parameter space, in which $M_{\smuL}<M_{\C1},M_{\N2}$ (the blue shaded region above the grey dashed line for $M_{\C1}=500 \gev$). However, $\C1/\N2$ decays via the left-smuon (and/or left-selectron) in this region are kinematically suppressed, and therefore, the $\geq 2\tau_h+\met$ search results shown remain effectively unchanged, even though they were obtained in Sec.~\ref{stau} with the assumption of $M_{\smuL}=3$ TeV. We see from Fig.~\ref{fig:gm21} that the $\gm2$ favoured parameter space at $1\sigma$ in the stau-NLSP scenario is severely constrained by the latest 13 TeV LHC results in the $\geq 2\tau_h+\met$ channel (red solid line). This conclusion is not modified under the variation of the relevant MSSM parameters, namely, $\tan \beta$, the mass of $\tla$ and the mixing angle in the stau sector, as we discuss below.

In Fig.~\ref{fig:gm21}, we have shown the results for $\tan\beta=50$. For lower values of $\tan\beta$, we have checked that the LHC search constraints are not modified significantly. However, as shown in Eqs.~\ref{eq:gm21}--\ref{b}, the MSSM contribution to $\Delta a_\mu$ is proportional to $\tan \beta$. Therefore, for lower values of $\tan\beta$, the $\gm2$ favoured region shifts towards lower values of the $\C1$ mass (for a fixed $\smuL$ mass), which translates to smaller values of $\mu$ and $M_2$. It is thus clear that for $\tan \beta$ values smaller than $50$, the current LHC constraints on the $\gm2$ favoured region are even stronger. 

As discussed in Sec.~\ref{stau}, the variation in $M_1$ does not affect the LHC search constraints, and its impact on $\Delta a_\mu$ through the neutralino-smuon loop is also small in the $\tilde{\mu}_R$ decoupled scenario. Furthermore, as explained in Sec.~\ref{stau}, with a maximally mixed $\tla$, the LHC constraints are only slightly weaker compared to the right-stau NLSP case, while the constraints are very similar with a left-stau NLSP. Therefore, changing the stau mixing would not relax the constraints compared to the red solid line in Fig.~\ref{fig:gm21}. As far as the $\tla$ mass is concerned, within the assumptions of the above mass hierarchy, its impact should be minimal as well, unless we encounter degeneracies between the $\C1/\N2$ and $\tla$ mass values. 

Therefore, we can conclude that in the stau NLSP scenario, with the mass hierarchy, $M_{\smuL}>M_{\C1},M_{\N2}>M_{\tilde{\tau}_1}>M_{\neu}$, the current 13 TeV LHC constraints already disfavour most of the parameter space in which the $\gm2$ anomaly can be accommodated within $1\sigma$ through the chargino contribution.

\subsection{Scenario-2: selectron/smuon NLSP}
\label{sec:five_two}
\begin{figure}[htb!]
\centering
\includegraphics[scale=0.6]{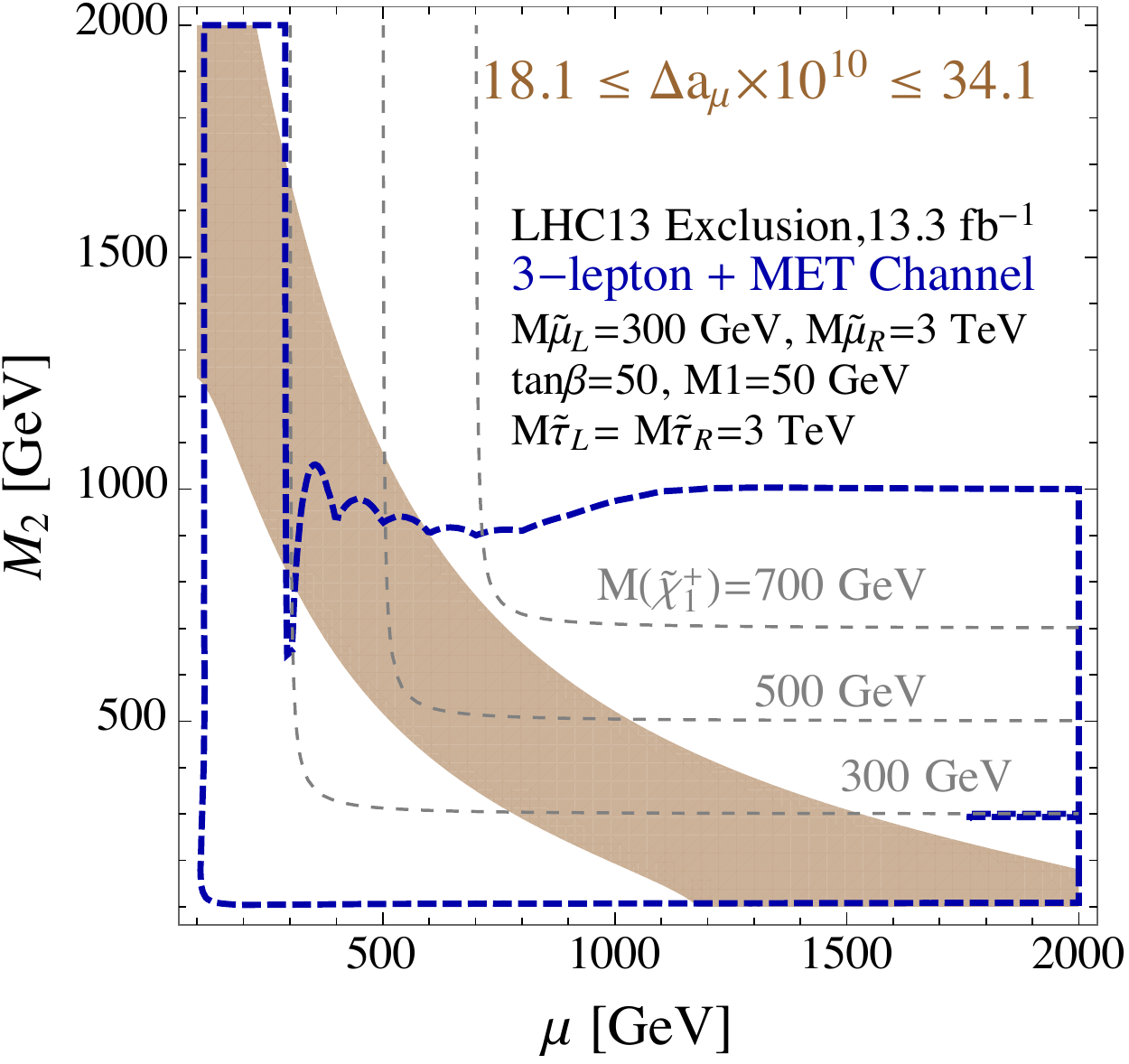}
\caption{\small\sl Current $95\%$ C.L.  LHC constraints on the $\gm2$ favoured parameter region in the $\mu-M_2$ plane, for the mass hierarchy $M_{\tilde{\tau}_1}>M_{\C1},M_{\N2}>M_{\smuL}, M_{\tilde{e}_L}>M_{\neu}$. In the brown shaded region, the $\gm2$ anomaly can be explained at $1\sigma$, with the choice $M_{\smuL}=300 \gev$. The values of other relevant MSSM parameters are fixed as in Fig.~\ref{fig:trilepton_all}.}
\label{fig:gm22}
\end{figure}

We now consider the second mass hierarchy: $M_{\tilde{\tau}_1}>M_{\C1},M_{\N2}>M_{\smuL}, M_{\tilde{e}_L}>M_{\neu}$. Since the $\smuL$ is the NLSP in this scenario, for our discussion, we choose the lowest value of $M_{\smuL}$ allowed by the LHC8 search for left-smuons described in Sec.~\ref{smuon}. In Fig.~\ref{fig:gm22}, we show the $\gm2$ favoured region at $1\sigma$ with $M_{\smuL}=300 \gev$ (brown shaded area), with $\tan \beta=50$. As we can see from this figure, in a substantial region of the $\mu-M_2$ plane, $M_{\C1}$ is lower than 300 GeV, which does not satisfy the hierarchy assumption of $M_{\C1},M_{\N2}>M_{\smuL}, M_{\tilde{e}_L}$. However, the $3\ell+\met$ search results shown in this figure (blue dashed line) were obtained with $M_{\smuL}=300 \gev$ in Sec.~\ref{trilepton}, and therefore the constraints shown remain valid in the entire region of the parameter space. In the region with $M_{\C1},M_{\N2}<M_{\smuL}, M_{\tilde{e}_L}$, the trilepton events originate from $\C1/\N2$ decays to $W^\pm/Z$ bosons. As we see from Fig.~\ref{fig:gm22}, even though 
the trilepton search is very powerful and strongly excludes most of the $1\sigma$ favoured parameter space for $\gm2$ with this mass hierarchy, in the Higgsino-like $\C1/\N2$ region, we can find a small window allowed by the current LHC search constraints. As discussed earlier, this is because of the reduced coupling of a left-smuon or left-selectron to a $\C1/\N2$, when the latter is mostly a Higgsino-like state. In the currently allowed region, the Higgsino mass parameter ($\mu$) is found to be around $500\gev$, with the wino-mass parameter ($M_2$) taking values higher than about 1 TeV, and the chargino mass falls in the range $300 \gev < M_{\C1}, M_{\N2} < 500 \gev$.

\subsection{Scenario-3: all three generation sleptons lighter than chargino}
\label{sec:five_three}
\begin{figure}[htb!]
\centering
\includegraphics[scale=0.6]{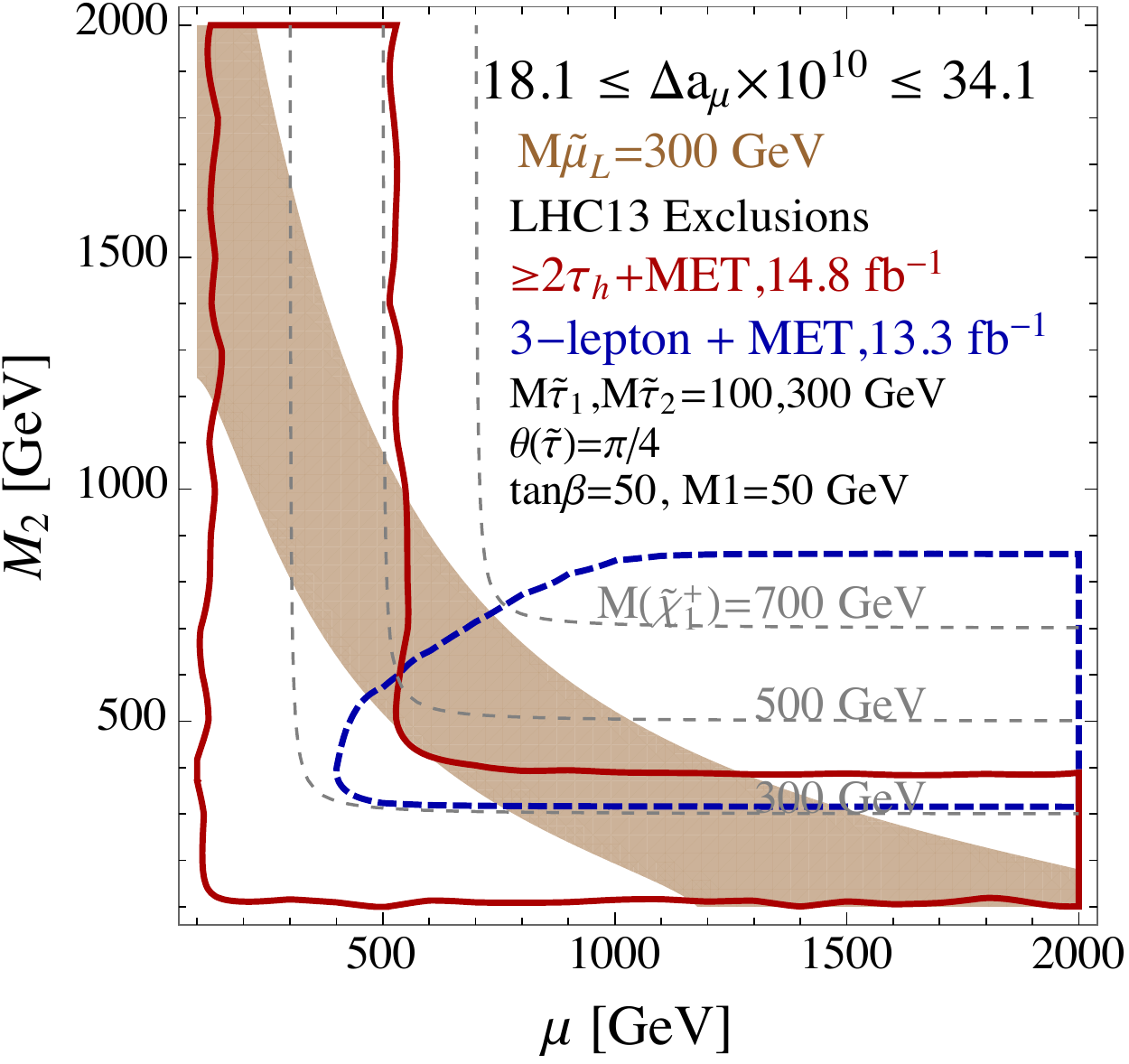}
\caption{\small\sl Current $95\%$ C.L. LHC constraints on the $\gm2$ favoured parameter region in the $\mu-M_2$ plane, for the mass hierarchy $M_{\C1},M_{\N2}>M_{\tilde{\tau}_1},M_{\smuL,\tilde{e}_L}>M_{\neu}$. In the brown shaded region, the $\gm2$ anomaly can be explained at $1\sigma$, with the choice $M_{\smuL}=300 \gev$. The values of other relevant MSSM parameters are fixed as in Fig.~\ref{fig:stau_smuon}.}
\label{fig:gm23}
\end{figure}

The third mass hierarchy considered in this study is $M_{\C1},M_{\N2}>M_{\tilde{\tau}_1},M_{\smuL,\tilde{e}_L}>M_{\neu}$. In Fig.~\ref{fig:gm23}, as in the previous case above, we show the $\gm2$ favoured region at $1\sigma$ for the left-smuon mass value of $M_{\smuL}=300 \gev$ (brown shaded), with $\tan\beta=50$. For discussing the LHC constraints, we choose, as in Fig.~\ref{fig:stau_smuon}, $M_{\tla}=100 \gev, M_{\tlb}=300 \gev$ and $\theta_{\tilde{\tau}}=\pi/4$, with $\tan \beta=50$ and $M_1=50 \gev$ (this choice of parameter values is motivated by the abundance for relic $\neu$ DM). Thus, both the $\smuL$ and the $\tla$ masses are fixed close to their current lower limits (for the case of $\tla$, the strongest bound from the LEP experiments is about $90 \gev$~\cite{Abbiendi:2003ji, SM}). Once again, as in the previous scenario, though taking $M_{\smuL}=300 \gev$ may violate the hierarchy of  $M_{\C1},M_{\N2}>M_{\smuL}$, the constraints shown in Fig.~\ref{fig:gm23}, namely that from the $\geq 2\tau_h+\met$ (red solid line) and $3\ell+\met$ (blue dashed line) searches were derived in Sec.~\ref{sec:sc3} for the same value of $M_{\smuL}$, and therefore they remain valid in the entire parameter space.

As discussed in Sec.~\ref{sec:sc3}, both the $\geq 2\tau_h+\met$ and the $3\ell+\met$ searches provide important complementary constraints for this mass hierarchy, and taken together, they exclude bulk of the $1\sigma$ favoured parameter region for $\gm2$, with $M_{\smuL}=300 \gev$. A small window remains allowed when $M_2 \sim \mu$ and $500 \gev < M_{\C1}, M_{\N2} < 700 \gev$, where both the searches lose their power. If $\smuL$ is made heavier, we will approach the stau NLSP scenario discussed above, and the overall constraints on the parameter space of interest will continue to be strong. 
\subsection{Comments on the lower bound on left-smuon mass and its impact on $\gm2$}
\begin{figure}[htb!]
\centering
\includegraphics[scale=0.6]{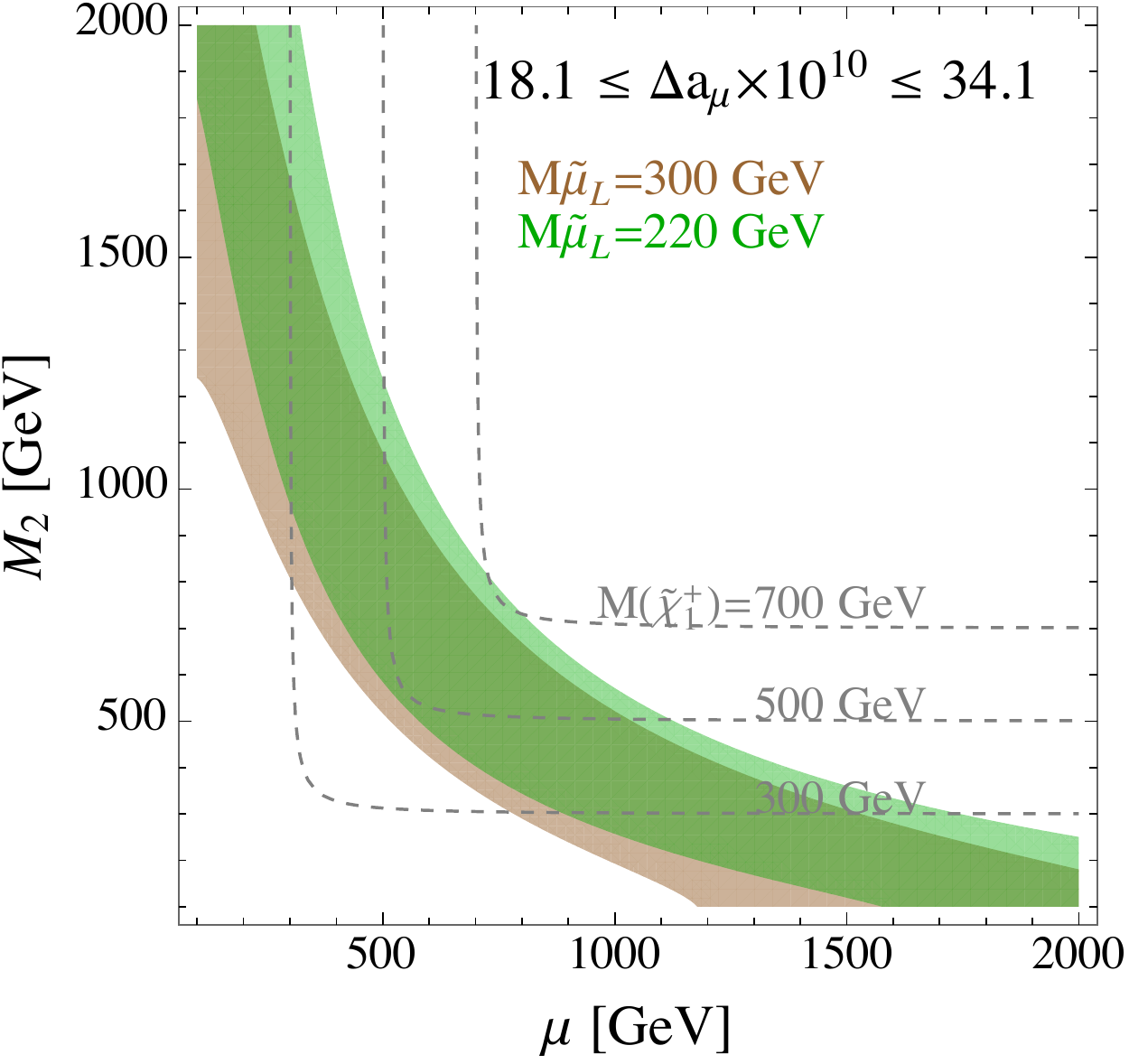}
\caption{\small\sl Parameter region in the $\mu-M_2$ plane in which the $\gm2$ anomaly can be explained within $1\sigma$: for $M_{\smuL}=300 \gev$ (brown shaded) and $M_{\smuL}=220 \gev$ (green shaded).}
\label{fig:gm24}
\end{figure}

In this subsection, we comment on the impact of having a weaker bound on $\smuL$ when the chargino is lighter than the left-smuon. As discussed in Sec.~\ref{smuon}, relaxing the 8 TeV LHC lower bound on the left-smuon mass from 300 GeV to around 220 GeV is possible only for a very light wino-like chargino, with $M_{\C1}$ smaller than $M_{\smuL}$. We show in Fig.~\ref{fig:gm24} the difference in the $1\sigma$ allowed regions for $\gm2$ for the two choices of $\smuL$ mass, $M_{\smuL}=300 \gev$ (brown shaded) and 220 GeV (green shaded). Focussing on the region below the grey dashed line along which $M_{\C1}<300 \gev$, we can see from this figure that the additional $\gm2$ favoured parameter space gained by lowering $M_{\smuL}$ is for $\mu>1500 \gev$ and $M_2<300 \gev$, where the lighter chargino is wino-like. However, from Figs.~\ref{fig:gm22} and \ref{fig:gm23}, we find that this additional parameter space is excluded by the current LHC search results on electroweak-inos. Therefore, the possible lowering of the $\smuL$ mass limit does not have any impact on our conclusions. 

We also note in passing that for very high $M_2$ and low $\mu$ values, the smuon mass bound cannot be relaxed, as the $\C1$ is Higgsino-like here, and thus the $\smuL$ cannot decay to a $\C1$. Therefore, the additional $\gm2$ allowed parameter space with $M_2>1700$ GeV and $\mu<300$ GeV in  Fig.~\ref{fig:gm24} is not viable either.

\section{Summary}
\label{sum}
In this study, we revisit the current LHC constraints on the $\gm2$ favoured parameter space in the MSSM, focussing on the scenario where the chargino-muon sneutrino loop contribution largely accounts for the discrepancy.  Since the chargino-muon sneutrino loop leads to the dominant contribution to $\gm2$ in most of the MSSM parameter space (except in the region in which the left- and right-scalar muons are light and there is a large mixing between them), it is crucial to thoroughly probe this scenario at the LHC.

The most relevant LHC searches in this context are that of the left-smuon and electroweak-inos, for which we include constraints from both the 8 TeV (with $20.3\fb^{-1}$ data) and 13 TeV (with upto $14.8\fb^{-1}$ data) LHC runs. In interpreting the LHC constraints from the above searches in the context of the $\gm2$, we organize our study by paying particular attention to the different possible mass hierarchies between the electroweak-inos and the three generations of sleptons. 

Since the LHC bounds on electroweak-inos become weaker if they decay via an intermediate stau or a tau sneutrino instead of the first two generation sleptons, we begin our study with the stau NLSP scenario, with the mass hierarchy $M_{\smuL}>M_{\C1},M_{\N2}>M_{\tilde{\tau}_1}>M_{\neu}$. For this mass hierarchy, we have carried out a detailed Monte Carlo analysis of the search for electroweak-inos in the $\geq 2\tau_h+\met$ channel. This is necessary because tau-jet identification and reconstruction efficiencies are sensitive to the event kinematics, and therefore to the mass differences among the supersymmetric particles involved. After carefully validating our MC simulation framework against the ATLAS results (as detailed in the Appendix), we translate the LHC8 and LHC13 bounds to the relevant Higgsino-wino mass ($\mu-M_2$) plane. Particular attention is paid to the possible dependence of the LHC constraints on the LSP mass and the mixing angle in the stau sector. It is observed that while the LSP mass does not play any significant role (unless we encounter degeneracies between the sparticle masses), the bounds are weaker for a left-stau or maximally mixed stau NLSP scenario, compared to a right-stau NLSP case. We note that the strongly mixed stau NLSP with a bino-like LSP is motivated by the DM relic density requirement (for $M_{\tilde{B}} \geq \mathcal{O}(30 \gev)$), or for accommodating the Galactic Centre gamma ray excess (for $M_{\tilde{B}} \sim 10 \gev$).

While the LHC constraints are found to be slightly weaker in a region where the $\C1/\N2$ are Higgsino-like, due to lower values of $\C1\N2$ production rates, this difference is reduced by the opening of new substantial production modes in the Higgsino-like region, such as that of $\C1 \tilde{\chi}_3^0$. With the left-smuon heavier than the $\C1/\N2$ for this mass hierarchy, we find that the $\gm2$ favoured region at $1\sigma$ is severely constrained even in the stau NLSP scenario, especially in the light of the recent 13 TeV data. Such strong constraints are obtained on setting $\tan \beta$ values as high as 50, and the LHC constraints on the $\gm2$ favoured parameter region become stronger for lower values of $\tan\beta$. 

When the left-smuon (and/or left-selectron) is the NLSP, we have the mass hierarchy $M_{\tilde{\tau}_1}>M_{\C1},M_{\N2}>M_{\smuL}, M_{\tilde{e}_L}>M_{\neu}$. In such a case, the LHC constraints come from the trilepton and $\met$ search channel, for which the limits on the electroweak-ino sector are generically stronger, and using the recently analyzed 13 TeV LHC data, ATLAS excludes $\C1/\N2$ masses of upto 1 TeV. We estimate the current constraints in the $\mu-M_2$ plane for this mass hierarchy as well, taking into account the modification to the $\C1\N2$ production cross-section and the corresponding decay BRs, in comparison to the simplified model based ATLAS study. With a left-smuon mass close to its current lower bound of around 300 GeV, we find that a small region of the $\gm2$ favoured parameter space at $1\sigma$ is still viable in view of the current LHC13 exclusion estimate for this mass hierarchy. In the allowed region, the Higgsino mass parameter ($\mu$) is found to be around $500\gev$, with the wino-mass parameter ($M_2$) taking values higher than about 1 TeV, and the chargino mass falls in the range $300 \gev < M_{\C1}, M_{\N2} < 500 \gev$. For such parameter values, the $\C1$ and $\N2$ are Higgsino-like, thereby reducing their BR to left-smuons, which leads to a weaker exclusion limit from the $3\ell+\met$ search.

The third possible mass hierarchy considered in this study is a scenario where all three generation sleptons are lighter than the chargino, i.e., $M_{\C1},M_{\N2}>M_{\tilde{\tau}_1},M_{\smuL,\tilde{e}_L}>M_{\neu}$. In this scenario, we find an interplay of the $\geq 2\tau_h+\met$ and the $3\ell+\met$ search channels, and the constraints ensuing from them in the $\mu-M_2$ plane are found to be largely complementary. For the Higgsino-like $\C1/\N2$ region ($\mu < < M_2$) the $\geq 2\tau_h+\met$ channel leads to stronger limits due to the higher decay rate of $\C1/\N2$ via an intermediate stau, whereas for the wino-like region ($\mu>>M_2$) the $3\ell+\met$ search leads to more stringent constraints. In the intermediate region, where the $\C1$ and the $\N2$ are nearly a maximal mixture of wino and Higgsino states ($\mu \sim M_2$), a small window of parameter space is found to be allowed based on the current constraints, in which the $\gm2$ anomaly can still be explained within $1\sigma$, with $500 \gev < M_{\C1}, M_{\N2} < 700 \gev$. Though individually both the multi-tau and multi-lepton searches lose power for such values of $\mu$ and $M_2$, it is plausible that a statistical combination of these two search channels would lead to stronger constraints. 

To sum up, on performing a thorough study of the constraints on the electroweak-ino sector parameters relevant for explaining the $\gm2$ anomaly via the dominant chargino-muon sneutrino loop in the MSSM, we find that with a bino LSP, and a slepton NLSP, the region of parameter space allowed by current LHC constraints is rather narrow. While for the stau NLSP scenario most of the viable parameter regions are already excluded, with a smuon (and/or selectron) NLSP, $300 \gev < M_{\C1}, M_{\N2} < 500 \gev$ are still allowed, when the lighter chargino is Higgsino-like. With all three generations of sleptons lighter than the chargino, the surviving region is found to be $500 \gev < M_{\C1}, M_{\N2} < 700 \gev$, with the lighter chargino being a maximal mixture of wino and Higgsino states.

We have not performed a detailed analysis for the fourth possible mass hierarchy of $M_{\smuL},M_{\tilde{\tau}_1}>M_{\C1},M_{\N2}>M_{\neu}$, in which all three generations of sleptons are heavier than the $\C1$ and $\N2$. In this scenario, the electroweak-inos would decay via SM gauge and Higgs bosons, and the LHC constraints become somewhat weaker compared to the above three scenarios. Therefore, it is expected that a part of the $\gm2$ favoured parameter space would be allowed, though the viable region shrinks compared to the above three cases due to the left-smuon being required to be heavier than the chargino.  Apart from this mass hierarchy, the $\gm2$ anomaly can also be accommodated by the contribution from the bino-left smuon-right smuon loop, with a large mixing between the left and right smuons, which in turn requires a large value of $\mu \tan \beta$; in addition, both the scalar muon states should be light.

In view of the fact that the upcoming FermiLab and J-PARC $\gm2$ experiments have the potential to confirm and possibly lead to an enhanced statistical significance of the $\gm2$ anomaly currently based on the BNL result, it is important at this point to determine the viability of well-motivated beyond standard model scenarios like the MSSM in explaining the current $3.3\sigma$ discrepancy. In this regard, as found in our study, the current LHC search results are already powerful enough to constrain large regions of the favoured parameter space. Future large statistics data sets from the LHC would not only consolidate the current constraints, but also probe regions of parameter space yet unexplored, which still remain promising from the point of view of explaining the $\gm2$ anomaly within the MSSM framework.

\section*{Note added}
After we submitted this paper for publication, new results from the 13 TeV LHC have been reported by the ATLAS and CMS collaborations which are relevant to the analyses presented here. The most important update comes from slepton search in the two leptons and missing transverse momentum final state, for which 13 TeV LHC results including $36.1{~\rm fb}^{-1}$ data is now available~\cite{multilepton_latest}. This improves the lower bound on $M_{\tilde{\mu}^\pm_L,\tilde{\mu}^\pm_R}$ from around $300$~GeV to $500$~GeV (for a common $\tilde{\mu}^\pm_{L,R}$ mass), with ${\rm BR}({\tilde{\mu}^\pm_{L,R}} \rightarrow \mu^\pm \neu) = 1$, and $M_{\neu}=0$ GeV. Furthermore, the update from the previously reported $13.3{~\rm fb}^{-1}$ to the current $36.1{~\rm fb}^{-1}$ results from 13 TeV LHC also modifies the multilepton search limits on chargino-neutralino pair production~\cite{multilepton_latest}. 

\begin{figure}[t]
\centering
\includegraphics[scale=0.5]{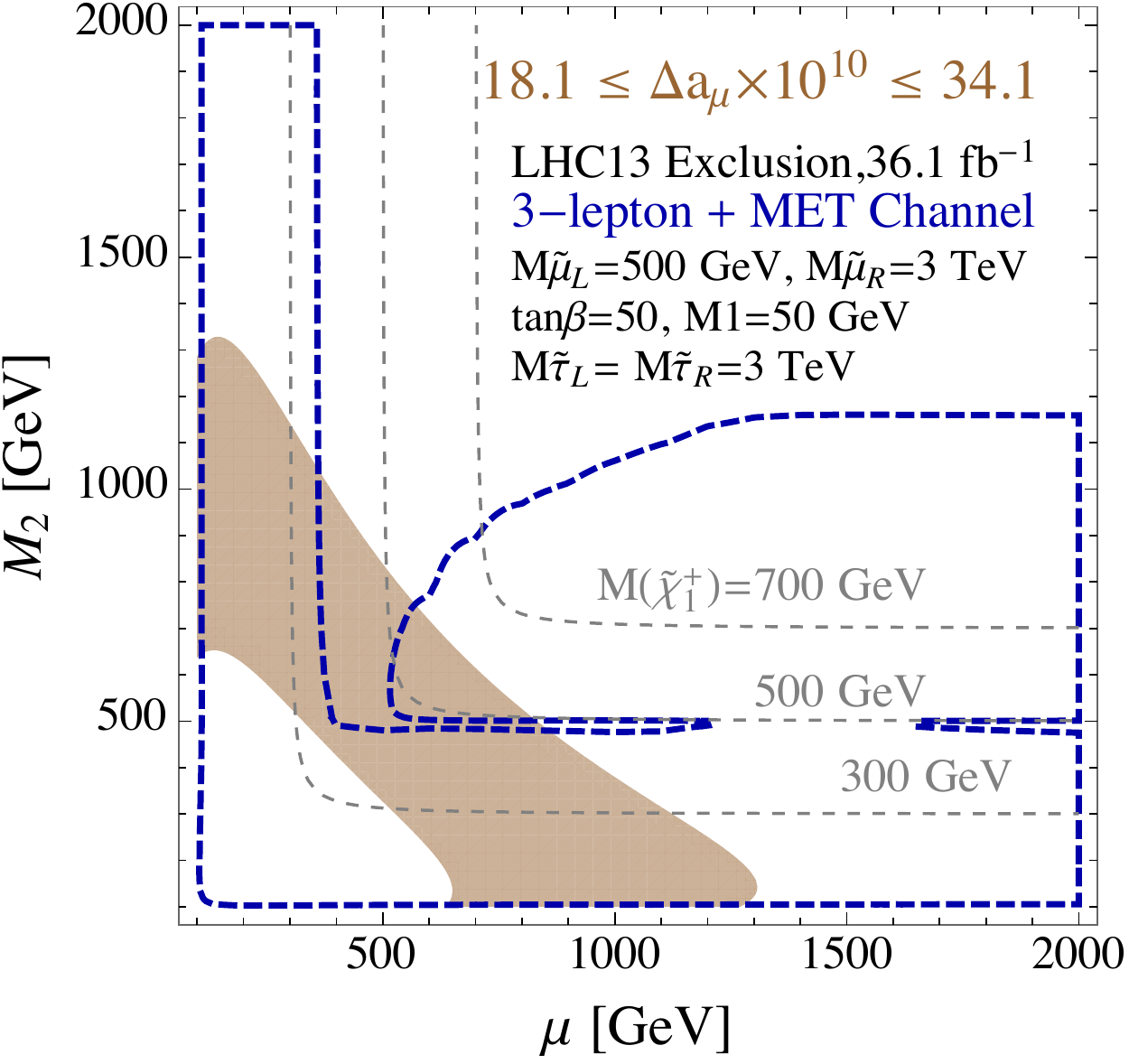}
\includegraphics[scale=0.5]{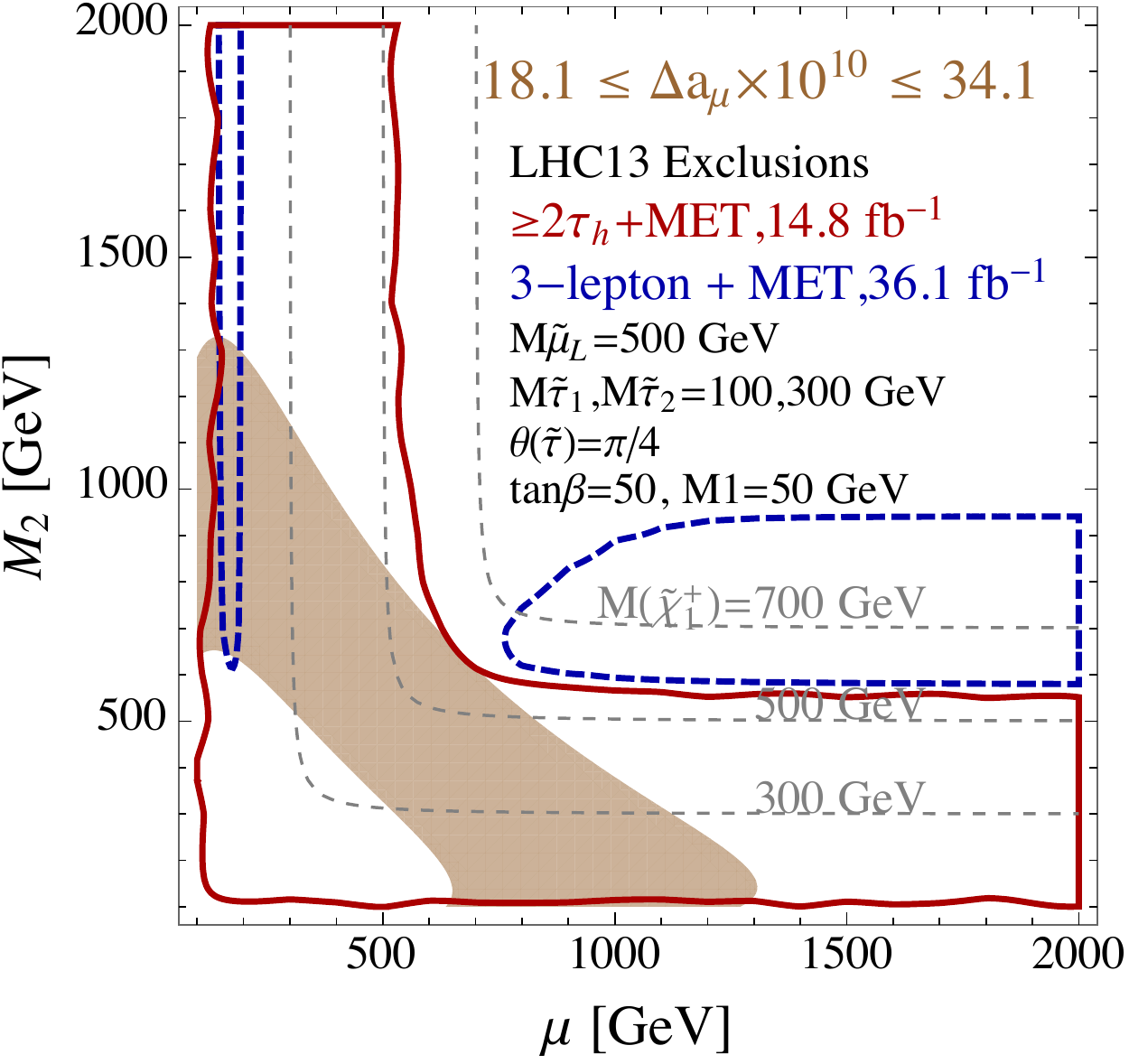}
\caption{\small\sl Updated LHC constraints on the $\gm2$ favoured parameter space using $36.1{~\rm fb}^{-1}$ of data in the multilepton$+\met$ search channel at the 13 TeV LHC~\cite{multilepton_latest}: for the selectron and/or smuon NLSP scenario (left column) and when all three generation sleptons are lighter than $\C1$ and $\N2$ (right column). See text in the Note added section for details.}
\label{fig:update_slepton}
\end{figure}

On including these new LHC results, we have obtained updated constraints in the $\gm2$ favoured parameter space in scenarios 2 and 3, which were studied in Sec~\ref{trilepton}, \ref{sec:sc3}, \ref{sec:five_two} and~\ref{sec:five_three}. In particular, we show the updated versions of Figs.~\ref{fig:gm22} and~\ref{fig:gm23} in Fig.~\ref{fig:update_slepton} (left and right columns respectively). To begin with, the $\gm2$ favoured region with $M_{\tilde{\mu}_L}=500$~GeV is now pushed to lower values of the $\C1$ mass. In the scenario with the selectron and/or smuon NLSP, in the currently allowed region, the Higgsino mass parameter ($\mu$) continues to be around $500\gev$, with the wino mass parameter ($M_2$) taking values between $500$~GeV and 1~TeV. We note that this region in the $\mu-M_2$ plane was covered by the multilepton search (blue dashed line in Fig.~\ref{fig:gm22}) with the previous bound of $300$ GeV on $M_{\tilde{\mu}_L}$. Since in this region $300 \gev < M_{\C1}, M_{\N2} < 500 \gev$, taking into account the updated constraint of $M_{\tilde{\mu}_L}>500$~GeV, the appropriate mass hierarchy to consider now is either scenario 1 with $M_{\neu} < M_{\tilde{\tau}_1} < M_{\C1},M_{\N2} < M_{\smuL}$ (in which case it is already excluded, see Fig.~\ref{fig:gm21}), or scenario 4 with
$M_{\neu} < M_{\C1},M_{\N2} < M_{\smuL},M_{\tilde{\tau}_1}$ (in which case it should still be allowed). With the stau sector parameter choices made in Fig.~\ref{fig:update_slepton}, left column, the currently allowed region belongs to scenario 4. 

In the scenario with all three generation sleptons lighter than $\C1$ and $\N2$, we find that with $M_{\tilde{\mu}_L}=500$~GeV, the $\gm2$ favoured parameter region is now covered almost entirely by the 13 TeV LHC search in the $\geq 2\tau_h+\met$ channel, as seen in Fig.~\ref{fig:update_slepton}, right column. We note that even though the $\geq 2\tau_h+\met$ channel search results have also been updated by the ATLAS collaboration including upto $36.1{~\rm fb}^{-1}$ data~\cite{tau_update}, the small improvements in the $\C1$ and $\N2$ mass limits do not impact our conclusions on the $\gm2$ favoured parameter space, which are already disfavoured in our analyses using the $14.8{~\rm fb}^{-1}$ dataset, following Ref.~\cite{ATLAS_tau_13}. 

\section*{Acknowledgments}
S.M. would like to thank Yasuhito Sakaki for help with {\tt Mathematica} graphics. K.H. and S.M. are supported in part by the U.S. Department of Energy under grant No.~DE-FG02-95ER40896 and in part by the PITT PACC. K.M. is supported by the China Scholarship Council, and the National Natural Science Foundation of China under Grant No. 11647018, and partially by the Project of Science and Technology Department of Shaanxi Province under Grant No. 15JK1150.

\section*{Appendix}
In this Appendix, we provide the details of our Monte Carlo (MC) simulation framework for the $\C1$ and $\N2$ searches in the $\geq 2\tau_h+\met$ channel. We follow the ATLAS analyses in this regard, details of which can be found in Ref.~\cite{ATLAS_tau_8} for the search using $20.3 \fb^{-1}$ of data from the 8 TeV LHC, and in Ref.~\cite{ATLAS_tau_13} for the search with  $14.8\fb^{-1}$ of data from the 13 TeV  run. 

For our MC simulation of a general MSSM scenario, we obtain the MSSM mass spectra from weak scale inputs of the soft SUSY breaking parameters using {\tt SUSPECT2}~\cite{Suspect}. We then compute the decay branching ratios (BR) of the supersymmetric particles using {\tt SDECAY}~\cite{Sdecay}, with the help of the {\tt SUSY-HIT} package~\cite{Susyhit}. The resulting mass spectra and decay BRs are used as inputs to {\tt MadGraph5}~\cite{MG5} for generating parton level events for electroweak-ino and stau pair production, which are then passed onto {\tt PYTHIA6}~\cite{Pythia} for including the effects of parton shower, hadronization and underlying events. We use {\tt DELPHES2}~\cite{Delphes} for including the effects of object reconstruction and detector resolution. Jets are reconstructed using the anti-$k_T$ algorithm~\cite{antikt} with radius parameter $R=0.4$ using {\tt FastJet}~\cite{Fastjet}. We employ the {\tt CTEQ6L1}~\cite{cteq,LHAPDF} parton distribution functions, and the factorization and renormalization scales have been fixed at the default event-by-event choice of {\tt MadGraph5}. 

\begin{figure}[t]
\centering
\includegraphics[scale=0.6]{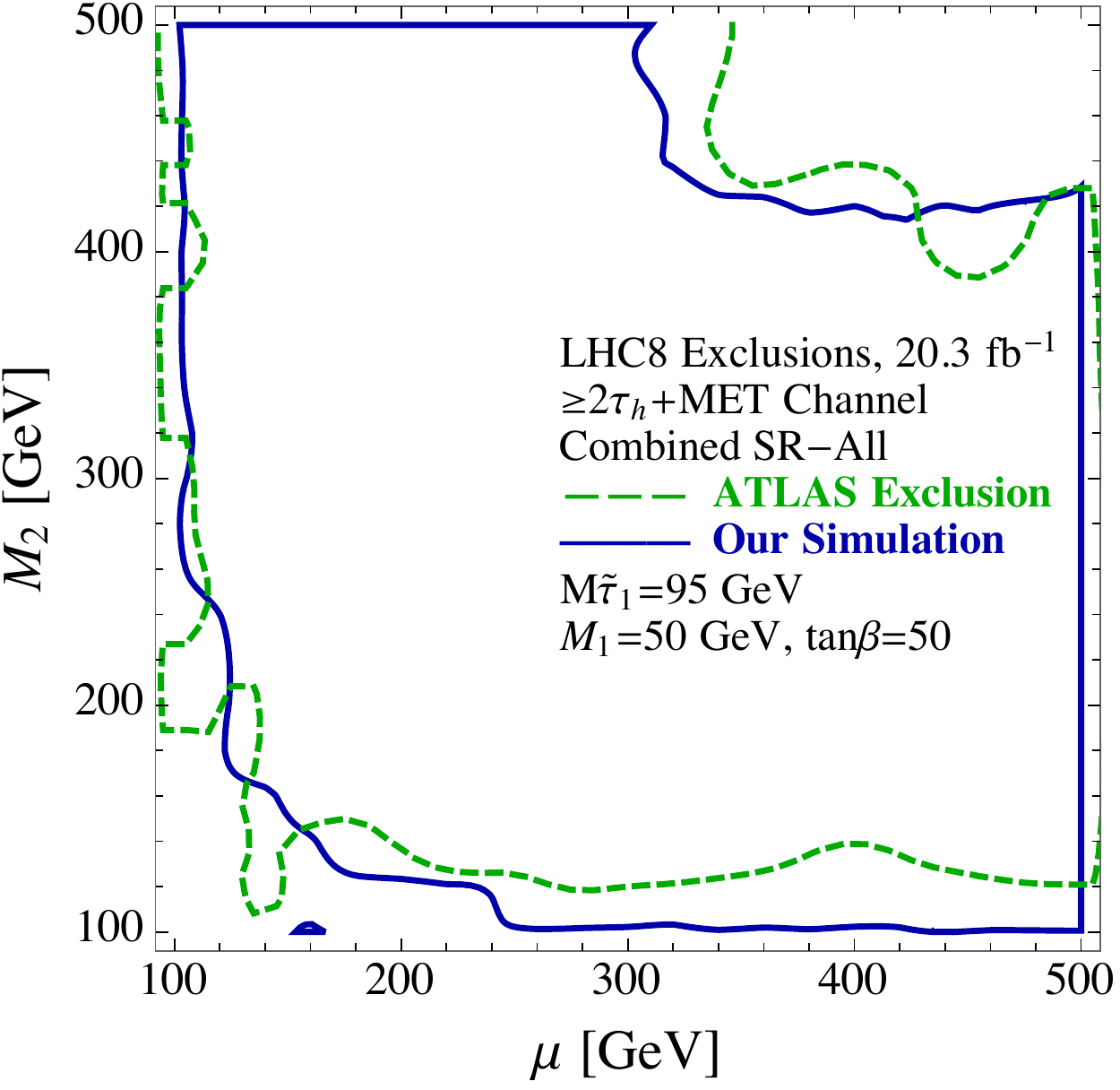}
\caption{\small\sl Comparison of the $95\%$ C.L. exclusion contours obtained by our MC simulation (blue solid contour) with the corresponding ATLAS pMSSM exclusion (green dashed contour) in the $\mu-M_2$ plane, using $20.3\fb^{-1}$ of data from the 8 TeV LHC, for the parameter choices $M_1=50 \gev, M_{\tla}=90 \gev$ and $\tan \beta=50$. For this analysis, the $\tla$ is taken to be a right-stau. See text for details.}
\label{fig:valid_8TeV}
\end{figure}
The 8 TeV ATLAS search is divided into four different signal regions, primarily targeting the $\C1\N2, \C1\C1$ and direct stau pair production processes. The 8 TeV results, however, are not sensitive to direct stau pair production exclusively~\cite{ATLAS_tau_8}, and most of the current sensitivity results from electroweak-ino pair production. The kinematic selection criteria used are as described in Ref.~\cite{ATLAS_tau_8}, see in particular the discussion in Sections 6 and 7, with the definition of different signal regions in Table 1.  For the three different MSSM reference points and the phenomenological MSSM (pMSSM) parameter space in the Higgsino-wino mass ($\mu-M_2$) plane studied by ATLAS, we find that our event yields for the signal process are somewhat larger than the numbers obtained by ATLAS. A very likely origin of this difference is the simple modelling of  tau-jet identification and reconstruction efficiencies in our MC simulation. It is, however, encouraging that we can approximately match the ATLAS results by multiplying our event yields with a constant fudge factor. After including an average K-factor in the range of $1.2-1.3$ to take into account the effects of higher order corrections to our leading order cross-section computation (we compute the next-to-leading order cross-sections using {\tt Prospino}~\cite{Prospino}), for the 8 TeV analysis, this fudge factor is found to be around $0.4$. In Fig.~\ref{fig:valid_8TeV}, we compare our $95\%$ C.L. exclusion (blue solid contour) with the ATLAS pMSSM exclusion (green dashed contour) in the $\mu-M_2$ plane, for the parameter choices $M_1=50 \gev, M_{\tla}=90 \gev$ and $\tan \beta=50$. For this analysis, the $\tla$ is taken to be a right-stau (see Figure 10 in Ref.~\cite{ATLAS_tau_8} for the corresponding ATLAS result). As we can see in this figure, our results are broadly in agreement with the ATLAS limits, after the above fudge factor is taken into account. In deriving the 8 TeV exclusion contours discussed in other sections of this study, we have adopted the same methodology.

\begin{figure}[t]
\centering
\includegraphics[scale=0.6]{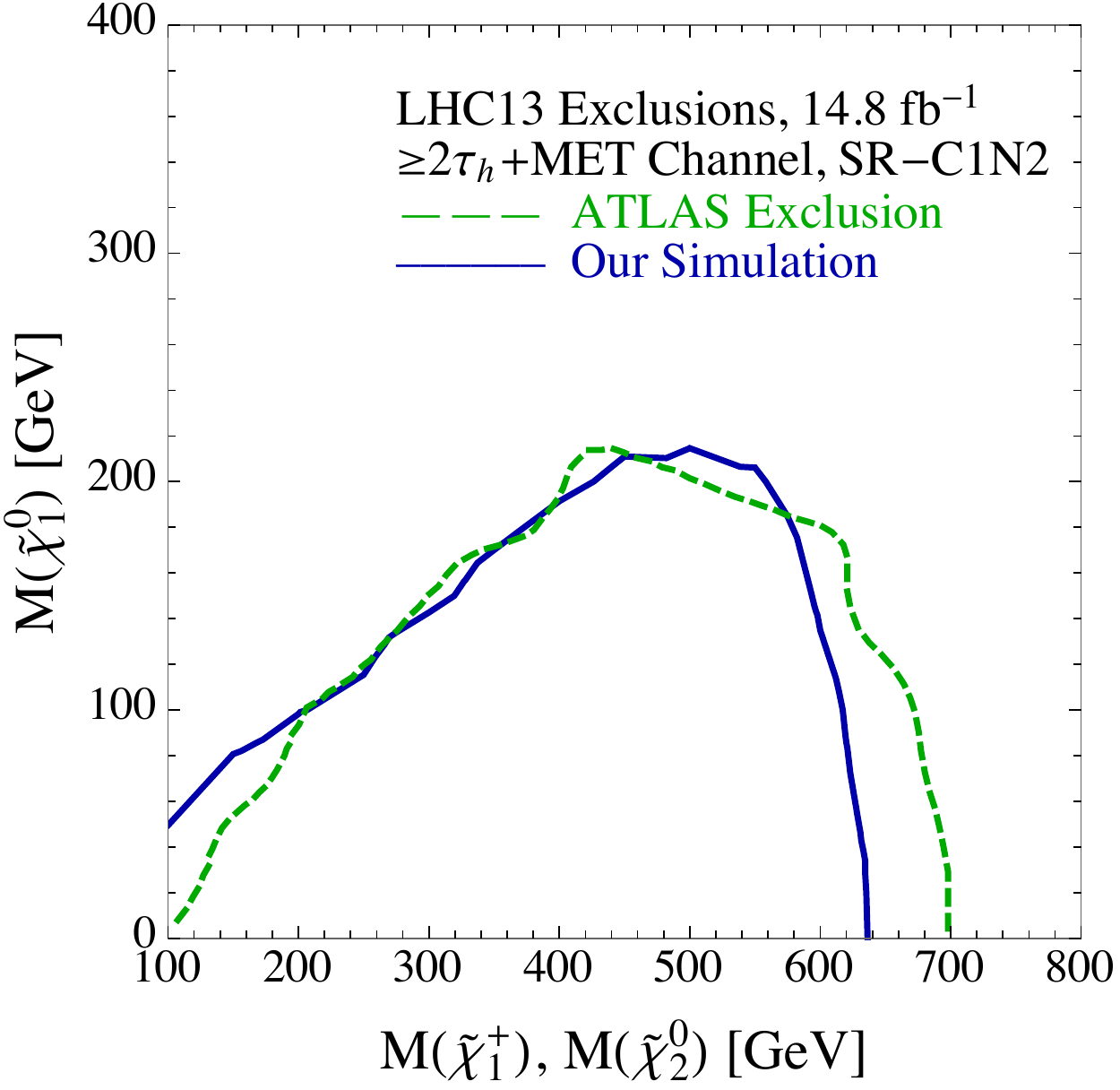}
\caption{\small\sl Comparison of the $95\%$ C.L. exclusion contours obtained by our MC simulation (blue solid contour) with the corresponding ATLAS exclusion (green dashed contour) in a simplified model study of $\C1\N2$ pair production, using $14.8\fb^{-1}$ of data from the 13 TeV LHC. The tau-slepton and tau-sneutrino masses are set to be $(M_{\C1}+M_{\neu})/2$, with $M_{\C1}=M_{\N2}$. See text for details.}
\label{fig:valid_13TeV}
\end{figure}
For the 13 TeV LHC study, we follow the recent ATLAS note~\cite{ATLAS_tau_13}, see in particular Sections 4 and 5 as well as Table 1 in Ref.~\cite{ATLAS_tau_13} for details on the object reconstruction, event selection and the definition of the different signal regions. For the 13 TeV analysis, so far, ATLAS has reported their results in terms of simplified models, and the corresponding interpretations for the pMSSM parameter space is not yet available. We therefore compare our results with the ATLAS limits on $\C1\N2$ pair production, where the $\C1$ and $\N2$ decay via an intermediate tau-slepton and tau-sneutrino of mass $(M_{\C1}+M_{\neu})/2$, with $M_{\C1}=M_{\N2}$. We compare the $95\%$ C.L. exclusion contours based on our MC simulation (blue solid) with the corresponding ATLAS results (green dashed) in Fig.~\ref{fig:valid_13TeV}. As in the 8 TeV analysis, we found our event yields to be larger than the ATLAS numbers, and it was necessary to normalize our signal yields by a fudge factor of around $0.14$, after taking the K-factor correction into account. This fudge factor is somewhat smaller than the one required for the 8 TeV analysis. This is possibly due to the dependence of the tau reconstruction and identification efficiencies on the tau-jet kinematics, which can be quite different depending upon the centre of mass energy, and the difference in the kinematic selection criteria adopted. Once again, after taking the fudge factor into account, the overall agreement of our MC simulation with the ATLAS results is found to be reasonably good, especially in the intermediate $\C1/\N2$ mass region, with our bounds being slightly weaker in the high mass region, and slightly stronger in the lower mass region~\footnote{Without taking into account the fudge factor, the obtained constraints from the multi-tau and missing transverse momentum channel would be stronger. However, since the $\gm2$ favoured 
parameter region where this search channel is important is already disfavoured in our 
analyses even with the fudge factor, our conclusions would remain the same.}.

%%%%%%%%%%%%%%%%%%%%%%%%%%%%%%%%%%%%%%%%%%%%

 \end{document}